\documentclass[fleqn,usenatbib]{mnras}

\usepackage{newtxtext,newtxmath}

\usepackage[T1]{fontenc}
\usepackage{ae,aecompl}
\newcommand{\vgrad}{\frac{{\rm d} v}{{\rm d}\chi}}
\usepackage{longtable}
\usepackage{supertabular}

\usepackage{graphicx}	
\usepackage{amsmath}	
\usepackage{amssymb}	
\newcommand{\kms}{{\rm kms}^{-1}}
\newcommand{\vrav}{$<v_r> = -109.0\pm1.6$~kms$^{-1}$}
\newcommand{\sigav}{$\sigma_{vr} = 7.8^{+1.7}_{-1.5}$~kms$^{-1}$}
\newcommand{\vg}{$\frac{{\rm d}v}{{\rm d}\chi} = -0.5\pm0.4$~kms$^{-1}$ per arcmin }
\newcommand{\thb}{$\theta=33^{+33}_{-27}\deg$}

\title[Andromeda XIX -- a detailed study]{A detailed study of Andromeda XIX, an extreme local analogue of ultra diffuse galaxies}

\author[M. L. M. Collins et al.]{Michelle L. M. Collins$^{1}$\thanks{E-mail: m.collins@surrey.ac.uk (MLMC)},
Erik J. Tollerud$^{2}$,
R. Michael Rich$^{3}$,
Rodrigo A. Ibata $^{4}$,
\newauthor Nicolas F. Martin $^{4,5}$, Scott C. Chapman$^{6}$, Karoline M. Gilbert$^{2,7}$, Janet Preston$^{1}$
\\
$^{1}$Physics Department, University of Surrey, Guildford, GU2 7XH, United Kingdon\\
$^{2}$Space Telescope Science Institute, 3700 San Martin Drive, Baltimore, MD 21218, USA\\
$^{3}$Department of Physics and Astronomy, University of California at Los Angeles, Los Angeles, CA 90095, USA\\
$^{4}$Observatoire astronomique de Strasbourg, Université de Strasbourg, CNRS, UMR 7550, 11 rue de l'Universit\'{e}, F-67000 Strasbourg, France\\
$^5$ Max-Planck-Institut für Astronomie, K\"{o}nigstuhl 17, D-69117 Heidelberg, Germany\\
$^6$Dalhousie University Dept. of Physics and Atmospheric Science
  Coburg Road Halifax, B3H1A6, Canada\\
$^7$Department of Physics \& Astronomy, Bloomberg Center for Physics and Astronomy, Johns Hopkins University, 3400 N. Charles Street, Baltimore, MD 21218  \\
}

\date{Accepted XXX. Received YYY; in original form ZZZ}

\pubyear{2018}

\begin{document}
\label{firstpage}
\pagerange{\pageref{firstpage}--\pageref{lastpage}}
\maketitle

\begin{abstract}
{With a central surface brightness of $\mu_0=29.3$ mag. per sq. arcsec, and half-light radius of $r_{\rm half}=3.1^{+0.9}_{-1.1}$~kpc, Andromeda XIX (And~XIX) is an extremely diffuse satellite of Andromeda. We present spectra for $\sim100$ red giant branch stars in this galaxy, plus 16 stars in a nearby stellar stream. With this exquisite dataset, we re-derive the properties of And~XIX, measuring a systemic velocity of \vrav\ and a velocity dispersion of \sigav (higher than derived in our previous work). We marginally detect a velocity gradient along the major axis of $\vgrad = -2.1\pm1.8~\kms$kpc$^{-1}$. We find  its mass-to-light ratio is higher than galaxies of comparable stellar mass ($[M/L]_{\rm half} = 278^{+146}_{-198}M_\odot/L_\odot$), but its dynamics place it in a halo with a similar total mass to these galaxies. This could suggest that And~XIX is a ``puffed up'' dwarf galaxy, whose properties have been altered by tidal processes, similar to its Milky Way counterpart, Antlia II. For the nearby stream, we measure  $v_r=-279.2\pm3.7~\kms$, and $\sigma_v=13.8^{+3.5}_{-2.6}~\kms$. We measure its metallicity, and find it to be more metal rich than And~XIX, implying that the two features are unrelated. Finally, And~XIX's dynamical and structural properties imply it is a local analogue to ultra diffuse galaxies (UDGs). Its complex dynamics suggest that the masses of distant UDGs measured from velocity dispersions alone should be carefully interpreted.}
\end{abstract}




\section{Introduction}
\label{sec:intro}

\begin{figure}
	\includegraphics[width=\columnwidth]{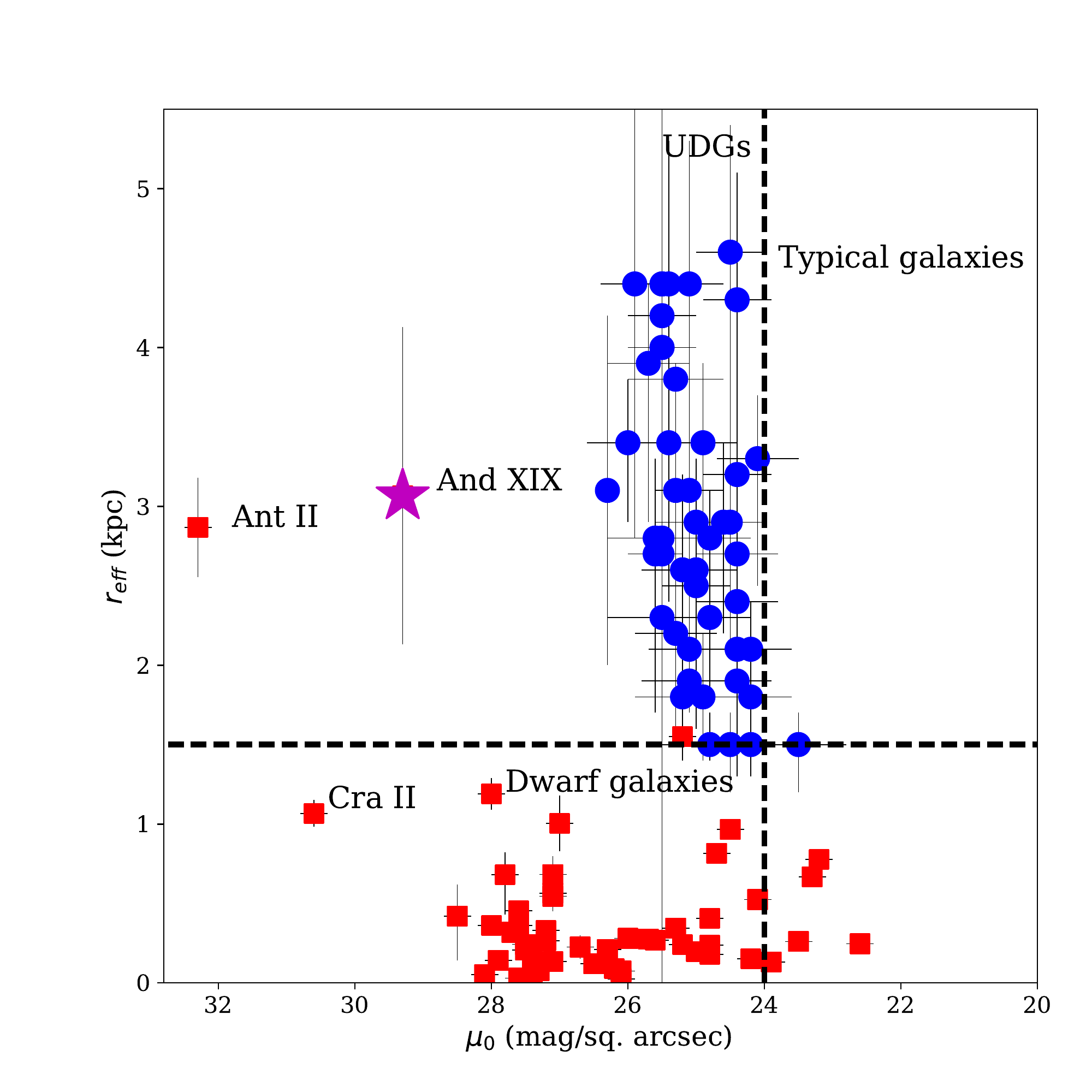}
    \caption{Central surface brightness ($\mu_0$) vs. effective radius ($r_{\rm eff}$ for Local Group dwarf galaxies (red squares) and Coma UDGs (blue circles). And~XIX is highlighted as a magenta star. While it is similar in size to the UDG galaxies, its surface brightness and stellar mass is significantly lower. }
    \label{fig:rhmu}
\end{figure}

In recent years, our view of the low surface brightness Universe has been revolutionised. Wide-field surveys like the Sloan Digital Sky Survey (SDSS, \citealt{york00}), Pan-STARRS1 \citep{chambers16}, and the Dark Energy Survey (DES, \citealt{abbott18}) have uncovered a population of extremely faint ($L\gtrsim100 L_\odot$), low surface brightness ($\mu_{V,0}<31$~mag arcsec$^{-2}$) dwarf galaxies within the Local Group. These faint systems further our understanding of galaxy formation in the low mass regime, and bring us closer to understanding where the lower limit for galaxy formation may lie. 
Further afield in the Coma cluster, advances in low surface brightness imaging have renewed interest in the study of low-surface brightness galaxies. Imaging with the Dragonfly telephoto array \citep{abraham14} revealed a vast population of diffuse ($\mu_{V,0}>24$~mag arcsec$^{-2}$), extended ($r_{\rm eff}>1.5$~kpc) systems (\citealt{vandokkum15a}, see fig.~\ref{fig:rhmu}). These `ultra diffuse galaxies' (UDGs) have sizes comparable to the Milky Way, but are orders of magnitudes fainter, comparable to dwarf galaxies. While similarly extreme objects have been known for some time \citep{binggeli84,impey88,bothun91,dalcanton97,conselice03}, these recent studies have shown that the UDG population in large clusters is vast, and extends across a range of different environments throughout the Universe \citep{koda15,merritt16,vanderburg17a}.

\begin{figure}
	\includegraphics[width=\columnwidth, angle=0]{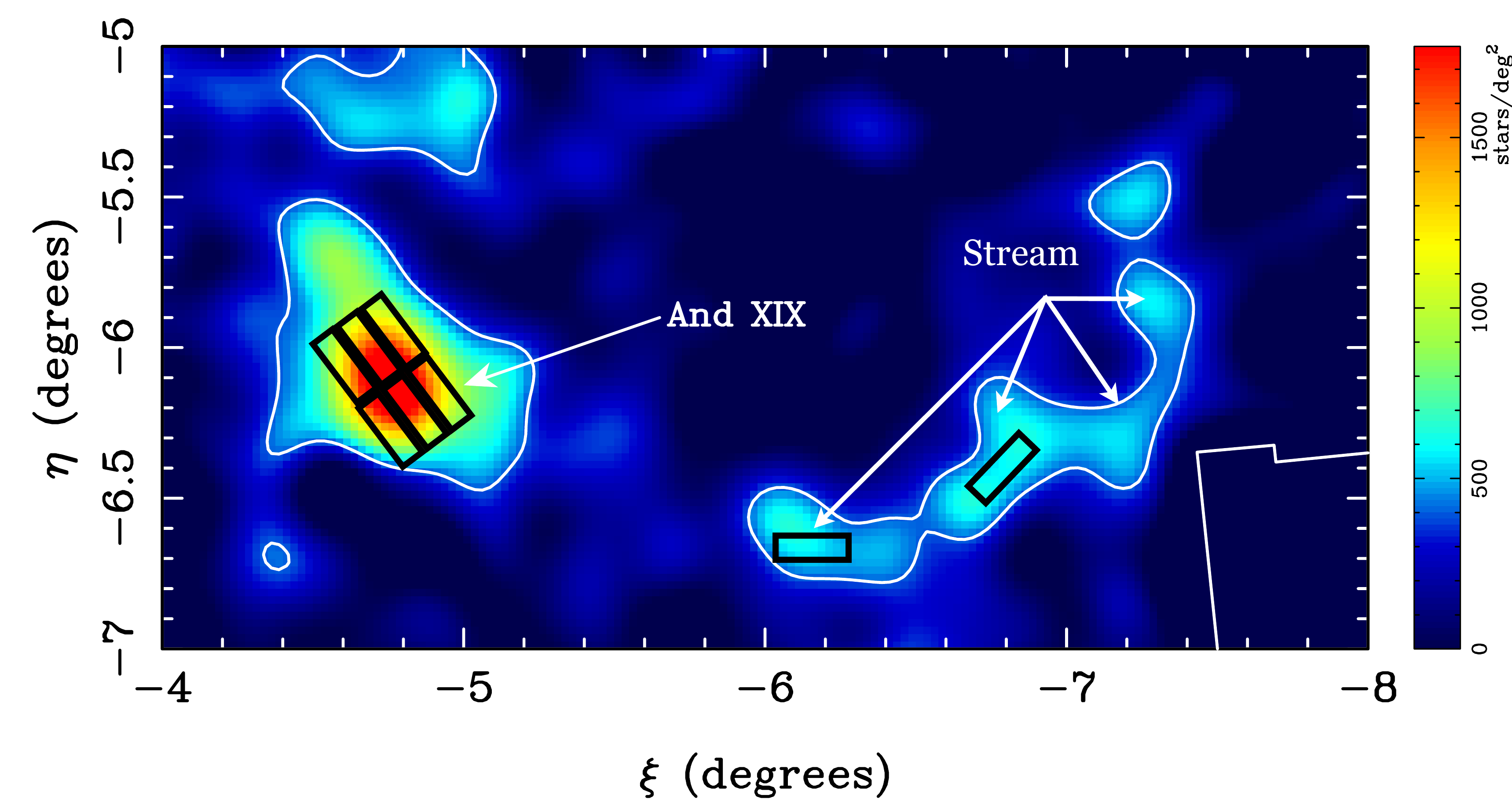}
    \caption{A surface density map of Andromeda XIX and its environs from the PAndAS survey. The map is created by counting the number of And~XIX like stars (based on position in the colour magnitude diagram) in pixels of size $0.5^\circ\times0.5^\circ$. This grid is smoothed with a Gaussian filter of width $\sigma = 0.1^\circ$. While the central $0.5^\circ$ appear smooth and round, the outskirts show evidence of stretching, possibly indicating tidal disruption. Black rectangles show positions of DEIMOS fields in And~XIX and the stream. The white polygons in the lower-right show the edge of the PAndAS survey.}
    \label{fig:a19pandas}
\end{figure}

Recently, two extremely low surface brightness, extended galaxies have been found in the Local Group. Using proper motions from the Gaia mission, \citet{torrealba18} uncovered the Antlia ~II (Ant~II) dwarf galaxy, which has a surface brightnesses of $\mu\gtrapprox32$~mag. sq. arcsec, and half-light radius of $r_{\rm half}\sim2.9$~kpc. The only other known galaxy with such properties is the Andromeda XIX (And~XIX) dwarf. And~XIX was first discovered by \citet{mcconnachie08}, and is one of the most extreme galaxies in terms of its size and surface brightness when compared to the UDGs discovered to date. It has a half-light radius of $r_{\rm half}=3065^{+935}_{-1065}$~pc and a central surface brightness of $\mu_{V,0}=29.3\pm0.4$ mag~arcsec$^{-2}$ \citep{martin16c}, placing it in a near-unique region of parameter space, accompanied only by Ant~II (fig.~\ref{fig:rhmu}). Located only $821^{+32}_{-148}$~kpc from us \citep{conn12b} in the halo of Andromeda (M31), And~XIX is close enough to resolve individual stars in both imaging and spectroscopy. Initial studies of its dynamics from $\sim25$ member stars  gave a velocity dispersion of $\sigma_v=4.7^{+1.6}_{-1.4}\kms$, and suggested that it inhabits a low mass dwarf halo, despite having an effective size comparable to the Milky Way \citep{collins13}. Imaging from the Pan-Andromeda Archaeological Survey (PAndAS) showed that And XIX may be tidally disrupting, as it shows elongated isophotes in its outskirts, and a nearby faint, stream-like feature  that may result from its tidal disruption (fig.~\ref{fig:a19pandas}, \citealt{mcconnachie08, martin16c}).

These diffuse galaxies (both within the Local Group and without) have presented the community with a puzzle: how do such large, diffuse galaxies form? As their surface brightnesses and stellar masses are most similar to dwarf galaxies, several scenarios link the UDGs to these systems (e.g. \citealt{conselice18}). These galaxies may form diffuse, within high spin halos (e.g. \citealt{amorisco16a}); or by the removal of gas reservoirs at early times through gas rich, star formation fueled outflows \citep{dicintio16}. Alternatively, some may be products of their environment, shaped by tidal stripping and harassment of more massive systems (e.g. \citealt{collins13,merritt16,carleton18,ogiya18,amorisco19a,mancerapina19,torrealba18}) For the nearby Ant-II, it has been shown that its properties can only be understood if it has experienced both extreme stellar feedback, and tidal stripping \citep{torrealba18}, implying there is more than one way to form a UDG. Alternatively, others have tried to link these galaxies to their similarly sized, more luminous counterparts, positing that they could be failed Milky Way galaxies \citep{vandokkum15b,vandokkum16}, although objects this massive must be rare, if they exist at all. 

To understand these various scenarios, one ideally needs information on the dynamics of the system. These can be used to measure the halo masses of UDGs, and to search for signs of tidal disruption (through streaming motions/rotation of stars). The majority of UDGs are located at large distances from the Milky Way, making detailed analysis challenging. Several studies have determined the halo masses of UDGs to see if they are more consistent with being massive, failed Milky Ways or diffuse dwarf galaxies. \citet{vandokkum19b} measured both the central velocity dispersion, and dispersion profile of one of the more extended Coma cluster UDGs, Dragonfly 44 (DF44), and found it is residing in a halo of $M\sim10^{11}M_\odot$, similar to that of the Large Magellanic Cloud (e.g. \citealt{penarrubia16,erkal19a}). Measuring the dynamics for globular clusters around UDGs in the Virgo cluster, \citet{toloba18} showed that these systems are quite dark matter dominated for their luminosity also (although one should be cautious when extrapolating mass from a velocity dispersion when it is not clear that the system is in equilibrium, e.g. \citealt{laporte18}). Other studies that use the specific frequency of globular cluster populations of UDGs as a proxy for mass find that these objects are most consistent with dwarf-massed dark halos (e.g. \citealt{beasley16,amorisco16b,prole19}). Studies of UDGs with HI gas also suggest their masses are most consistent with dwarf galaxies \citep{trujillo16a,leisman17a}, and constraints from weak-lensing paint the same picture \citep{sifon18}.

To better understand the possible formation channels for UDGs, we present detailed study of the kinematics and spectroscopically derived metallicities for approximately 100 red giant branch stars (RGBs) in the  curious And~XIX dwarf. We also present data from two fields within the stream-like structure to determine whether it is linked to And~XIX. Using this exquisite dataset, we assess the current properties of And~XIX, investigate the mass of its halo and discuss likely formation scenarios for this object. This paper is laid out as follows: we detail our photometric and spectroscopic observations in \S~\ref{sec:obs}; our methods and results are shown in \S~\ref{sec:props}; we discuss the possible origins for And~XIX, and the unusual features discovered both spectroscopically and in the imaging in \S~\ref{sec:disc}, and we conclude in \S~\ref{sec:conc}.

\section{Observations}
\label{sec:obs}

\subsection{Suprime-cam imaging of And XIX}
\label{sec:pobs}

\begin{figure}
	\includegraphics[width=\columnwidth]{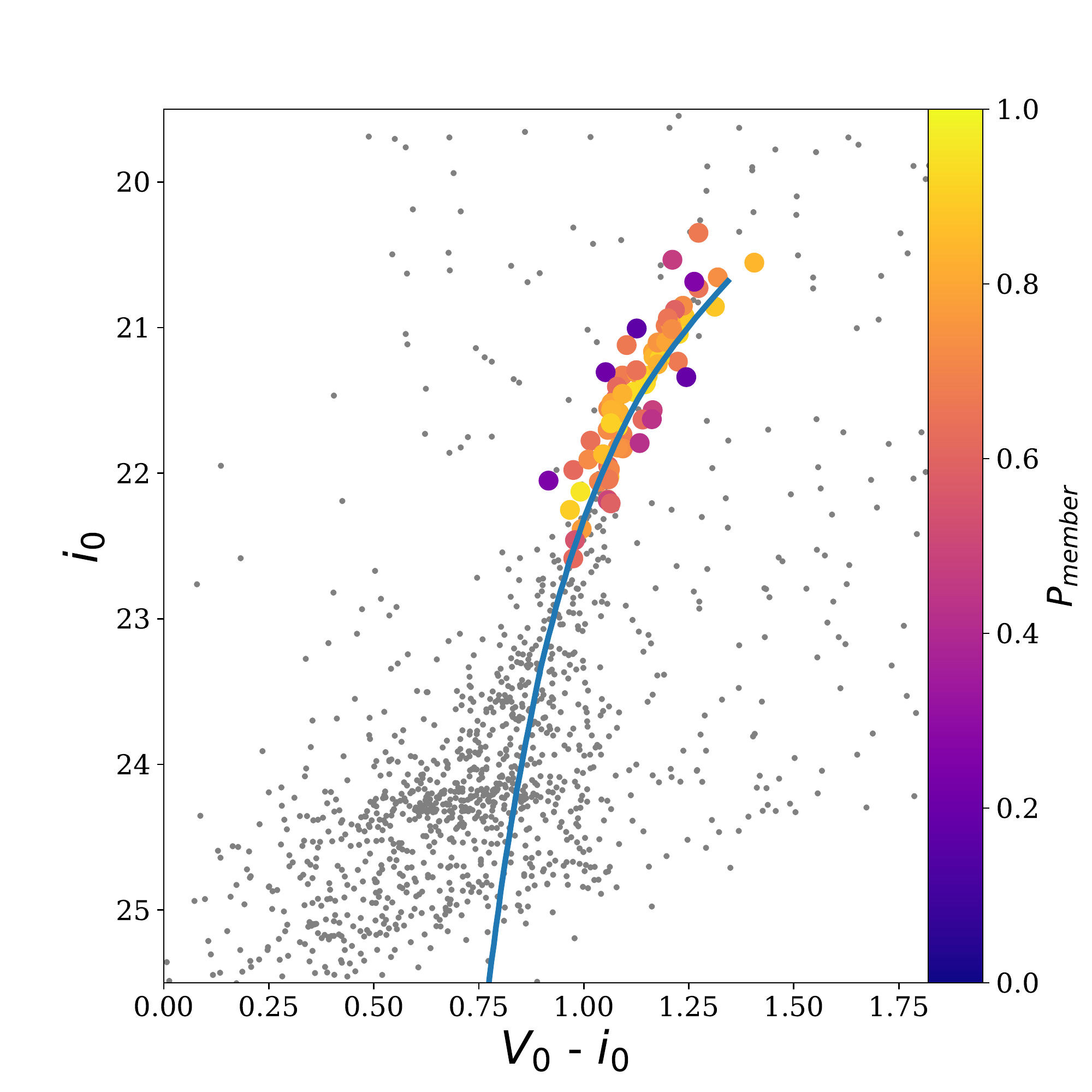}
    \caption{Suprime-cam CMD for And~XIX. Stars from our DEIMOS spectroscopy with $P_{\rm member}>0.1$ are shown, colour coded by their membership probability. We have overlaid an old, metal poor isochrone from the {\sc Parsec} stellar evolutionary models ([Fe/H]$=-1.8$, $[\alpha/$Fe]$=0.0$, age = 12~Gyr, shifted to a distance modulus of $m-M=24.57$ \citealt{bressan12,conn13}) that well represents the RGB of the dwarf galaxy }
    \label{fig:CMDs}
\end{figure}

We conducted deep imaging of And~XIX using Suprime-cam on the Subaru telescope on 24th August 2008 in photometric conditions, with sub-arcsecond seeing (average of $\sim0.7^{\prime\prime}$). Suprime-cam has a field of view of $27\times34$ arcmin, allowing coverage of $\sim 1$ half-light radii of And~XIX in this single pointing. We used the wide $V-$ (JC) and $i-$band (AB) filters, integrating for 3$\times400s$ and $9\times220s$ in each band respectively. These deep observations image the stellar populations in And~XIX to below the red clump and horizontal branch (see fig.~\ref{fig:CMDs}). The data were processed using the CASU pipeline for processing wide-field optical CCD data \citep{irwin01}. The images were debiased and trimmed before being flat-fielded and gain-corrected to a common internal system using master flats constructed from twilight sky observations. Catalogues were generated for every science image and used to refine their astrometric alignment. The images were then grouped for individual objects and passbands and stacked to form the final images based on the updated astrometry. A catalogue was then generated for each final stacked image, objects morphologically classified as stellar, non- stellar or noise like, and the $V-$ and $i-$band catalogue data merged. Owing to the large size of And~XIX on the sky ($r_{\rm half} = 14.2$ arcmin), these data are not wide-field enough to rederive structural properties of the dwarf galaxy. For these, we use the properties as derived from the PAndAS survey \citep{martin16c}.

\subsection{Spectroscopic observations of And XIX stars}
\label{sec:sobs}

For the dynamics and metallicities of individual And XIX stars, we employed the Deep Extragalactic Imaging Multi-Object Spectrograph (DEIMOS, \citealt{deep2,davis03,cooper12}) on the Keck II telescope. The DEIMOS field of view ($5^{\prime}\times16.7^{\prime}$) allows coverage of stars within roughly one half-light radius of And XIX per pointing ($r_{\rm half}=14.2^{+1.4}_{-3.6}$~arcmin, \citealt{martin16c}).  We used the 1200l/mm grating  (resolution$\sim0.33$\AA$^{-1}$), a central wavelength of 8000\AA, and the OG550 filter. This covers a spectral range from $\sim6000-9000$\AA, which includes the calcium triplet (Ca~II) absorption feature. These lines are used to determine stellar velocities and metallicities ([Fe/H]) for our targeted stars.

As And~XIX is both large on the sky and diffuse ($\mu_0=29.3$ mag arcsec$^{-2}$, \citealt{martin16c}, fig.~\ref{fig:rhmu}), a multi-year campaign was required to maximise the number of member stars observed. Initial results from 2 masks, presented in \citet{collins13}, measured velocities for only 24 members. These initial data implied a low velocity dispersion of $\sigma_v=4.7^{+1.6}_{-1.4}\kms$, suggesting And~XIX resided in a very low-mass halo. In order to understand this surprisingly low velocity dispersion, and unique structure of And~XIX,  a further 9 DEIMOS pointings were made between Sept 2012 and Sept 2016.  The details of each observation are summarised in table~\ref{tab:specobs}. Briefly, masks were observed for either 1 or 2 hours (split into 3 or $6\times20$~minute integrations). The final set of observations has provided  $136$ velocities for And~XIX stars (a total of 96 independent stars, 40 of which have repeat measurements). In our full dataset, there are 115 stars with repeat observations. These allow us to determine any night-to-night variations in our observations, and pin down systematic uncertainties in our velocity measurements (similar to \citealt{simon07}).

\subsubsection{Selecting targets for DEIMOS observation}
\label{sec:sobs:targets}

For And~XIX, stars were selected as targets for each of our DEIMOS observations using the Subaru Suprime-Cam imaging. For the stream feature, we used  imaging from the Pan-Andromeda Archaeological suvey \citep{mcconnachie18}. In both, we isolate the region of colour-magnitude space that the RGB of And~XIX is located in using its colour magnitude diagram (CMD, fig.~\ref{fig:CMDs}). We assigned a priority to each star on this sequence depending on its $i-$band magnitude. Stars lying directly on the RGB, with $20.1 < i_0 < 22.5$ were given a high priority (priority A), followed by stars on the RGB with $22.5 < i_0 < 23.5$ (priority B). The remainder of the masks were filled with stars in the field with $20.3<i_0 <22.5$ and $0.5<(V - i)_0<4$ (priority C). Then, we used the {\sc IRAF DSimulator} package to design our multi-object masks. For each mask $> 100$ stars are targeted. After reducing and analysing the data, we find between 10-20 And~XIX members per mask, an efficiency of $\sim15\%$. In our stream fields, we found 10 members in one field, and 6 in the second.

\begin{table*}
	\centering
	\caption{Details of And~XIX spectroscopic observations. A total of 136 And~XIX velocities were measured, for 96 independent stars (40 repeat measurements)}
	\label{tab:specobs}
	\begin{tabular}{lcccccccc}
		\hline
		Mask name & Date & RA & Dec  & Position angle (deg)& Exposure time ($s$) & No. targets & No. members \\
		\hline
		7A19a & 2011-Sep-25 &00:19:36.7  & +35:06:42 & 90 &  3600  & 107 &17\\
		7A19b & 2011-Sep-25 & 00:19:30.4 & +35:07:34 & 0 & 3600 & 103 & 7 \\
		8A19a & 2012-Sep-20 & 00:19:23.7 & +35:05:40 & 37 &  3600 & 77 & 13\\
		8A19b &  2012-Sep-20 & 00:19:41.6 & +35:03:32 & 217 &  3600 & 80 & 9 \\
		8A19c &  2012-Sep-21 & 00:19:15.4 & +34:56:26 &  37 & 3600 & 72 & 11 \\
		A19m1 &  2014-Sep-17 &00:19:49.4 & +35:06:50  & 40 &  7200 & 102 & 22\\
		A19m2 &  2014-Sep-17 & 00:19:09.9 & +34:57:17 &  40 & 7200 & 91 & 7 \\
		A19l1 &  2014-Sep-21 & 00:20:17.0 & +35:02:52 &  40 & 7200 & 98 & 17 \\
		A19l2 &  2014-Sep-21 & 00:18:50.8 & +35:00:11 &  40 & 7200 & 86 &  11\\
		A19r1 &  2014-Sep-22 & 00:19:38.7 & +35:11:12 &  40 & 7200 & 100 & 9 \\
		A19r2 &  2016-Sep-04 & 00:19:30.3 & +34:57:41 &  40 & 7200 & 88 & 13\\
		Stream 1 & 2014-Sep-22 & 00:10:00.9 & +34:40:12 &  0 & 7200 & 104 &  10\\
		Stream 2 & 2014-Sep-22 & 00:13:13.9 & +35:01:32 &  45 & 7200 & 90 &  6\\
		\hline
	\end{tabular}
\end{table*}

\subsubsection{Data reduction}
\label{sec:sobs:reduc}

The data were reduced using two custom DEIMOS pipelines. The first was developed by \citet{ibata11}, and is described in detail by \citet{collins13}. The second is based on the {\sc Spec2D} pipeline \citep{cooper12,newman13}. These pipelines are the standard machinery for the PAndAS and Spectroscopic and Photometric Landscape of the Andromeda  Stellar Halo (SPLASH, e.g. \citealt{gilbert09}) teams respectively, and have been used to analyse DEIMOS observations of M31 dSphs in the past (e.g. \citealt{kalirai10,collins10,collins11a,collins13,collins15,tollerud12,tollerud13}). By using both pipelines in our analysis, we test their consistency. The techniques used in each pipeline are broadly similar, and are described below.

First, our raw images are reduced to one-dimensional spectra. We then measure the line-of-sight velocities for our stars ($v_{r, i}$) by cross-correlating their 1D spectra with the spectra of known radial velocity standard stars ({\sc Spec2D}), or stellar templates \citep{ibata11}. The velocities and uncertainties ($\delta_{vr, i}$) on these are generated in 2 different ways. For the \citet{ibata11} pipeline, we use a Markov Chain Monte Carlo procedure (MCMC) where a template Ca~II spectrum is cross-correlated with non-resampled data. This generates a most-likely velocity for each star, and an uncertainty based on the posterior distribution function. For the {\sc Spec2D} pipeline, we use a Monte Carlo procedure, wherein we re-simulate each spectrum with added noise, representative of the per-pixel variance. We then re-determine the velocity for this spectrum, and repeat the process 1000 times. The final velocity and uncertainty are then the mean and variance from these re-simulations \citep{simon07}. By using these two pipelines, we have been able to determine that they produce consistent results for the velocities of RGB stars in M31 (discussed in Appendix~\ref{sec:pipeline}). 

For both pipelines, we include a systematic uncertainty for our velocities that has been derived from DEIMOS observations of our 115 stars with repeat measurements. We find this systematic floor in our dataset to be $\sigma_{\rm DEIMOS}=3.2\kms$ (see Appendix \ref{sec:pipeline}), slightly higher than the values derived in previous studies of $\sigma_{\rm DEIMOS}=2.2$~kms$^{-1}$ \citep{simon07,tollerud12}. This could be due to the slightly larger jitter present in RGB stars. A similar value was measured by \citet{martin14b} in their study of 3 Andromeda satellites. We add this uncertainty in quadrature to our measured uncertainty. Finally, as DEIMOS is a slit spectrograph, we also correct for any small shifts in the wavelength solution that can occur from mis-centring of stars within the slits themselves. We do this using strong telluric lines to refine the wavelength solution, as outlined in, e.g., \citet{tollerud12}.

For metallicity measurements, we restrict our analysis to stars with $S/N>5$\AA$^{-1}$. Stars with $S/N$ this low will still have large uncertainties on their measured [Fe/H], but are less likely to produce spurious [Fe/H] measurements from misidentification of skylines as Ca~II lines. We prepare each spectrum for analysis as follows. We normalise the continuum by applying a median filter, which approximates the continuum with a smoothed fit. We divide the original spectrum by this continuum fit, resulting in a flat, normalised spectrum. We next fit the 3 Ca~II lines and the continuum simultaneously. We convert the areas from these fits into equivalent widths. We then use the sum of the second and third equivalent widths for our [Fe/H] calculation, following the procedure of \citet{starkenburg10}. To determine the uncertainties in our metallicity estimates, we use the $1\sigma$ uncertainties of the equivalent width measurements to determine an upper and lower bound on [Fe/H].

\section{The properties of And XIX}
\label{sec:props}

\begin{figure*}
     \includegraphics[angle=0,width=\columnwidth]{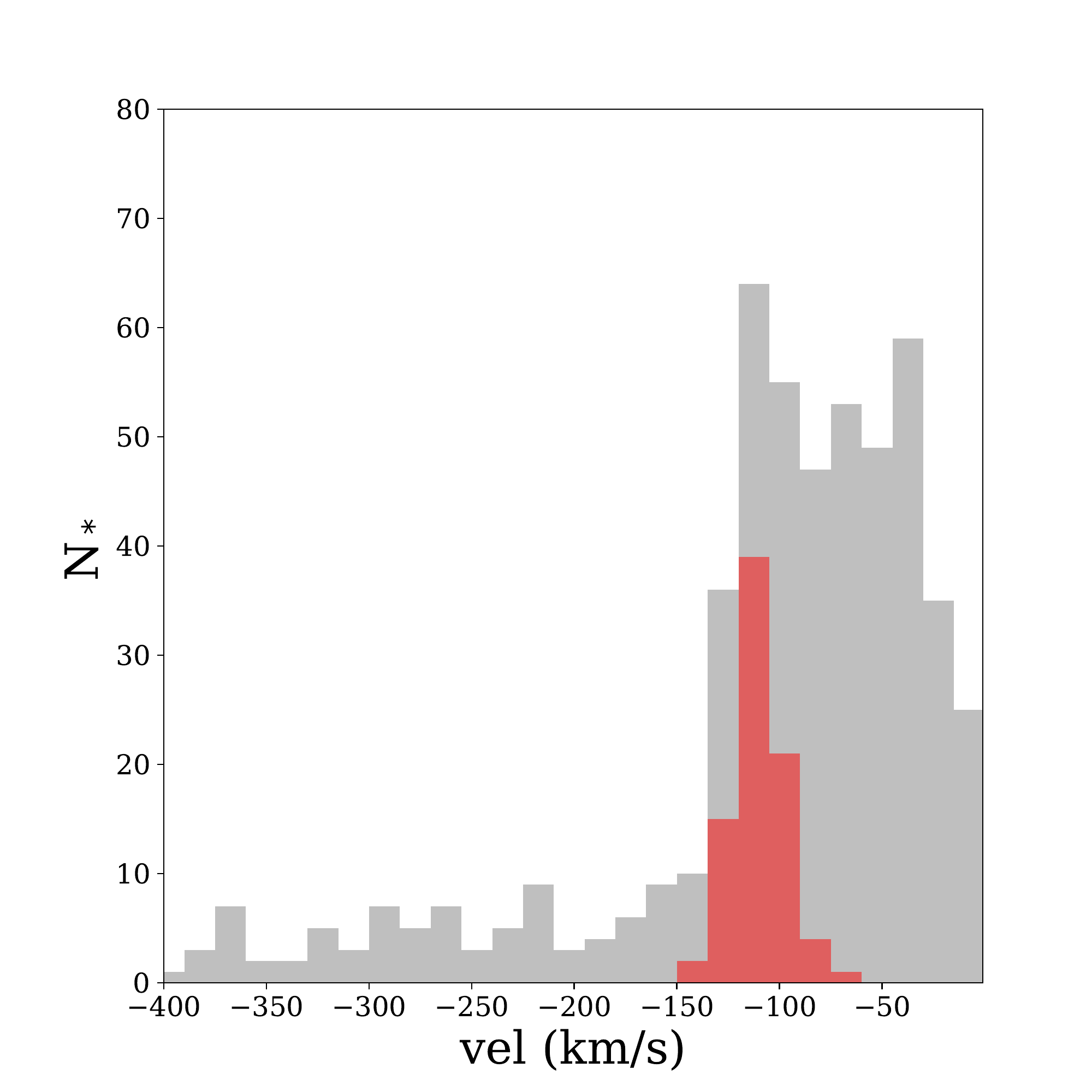}
     \includegraphics[angle=0,width=\columnwidth]{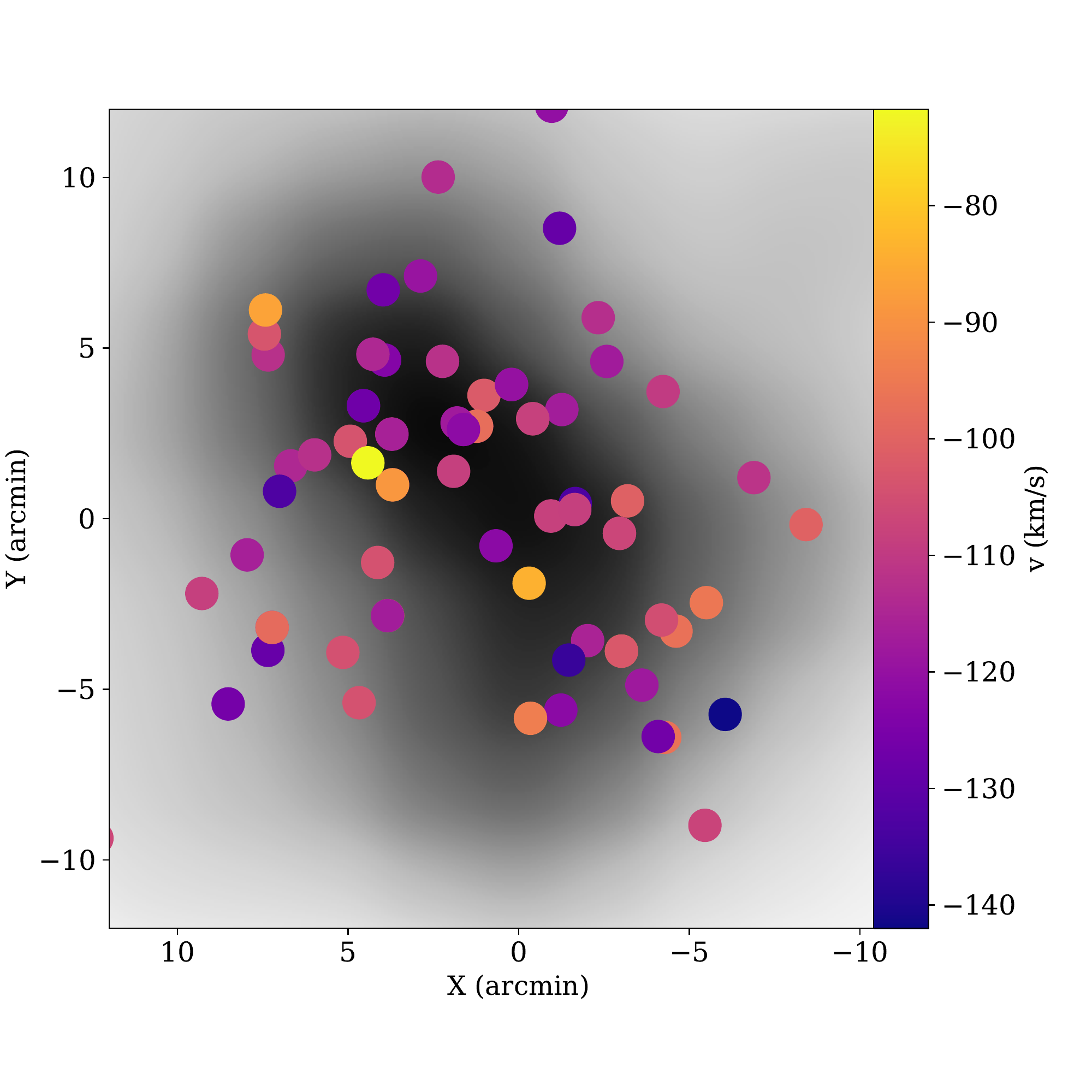}
     \includegraphics[angle=0,width=\columnwidth]{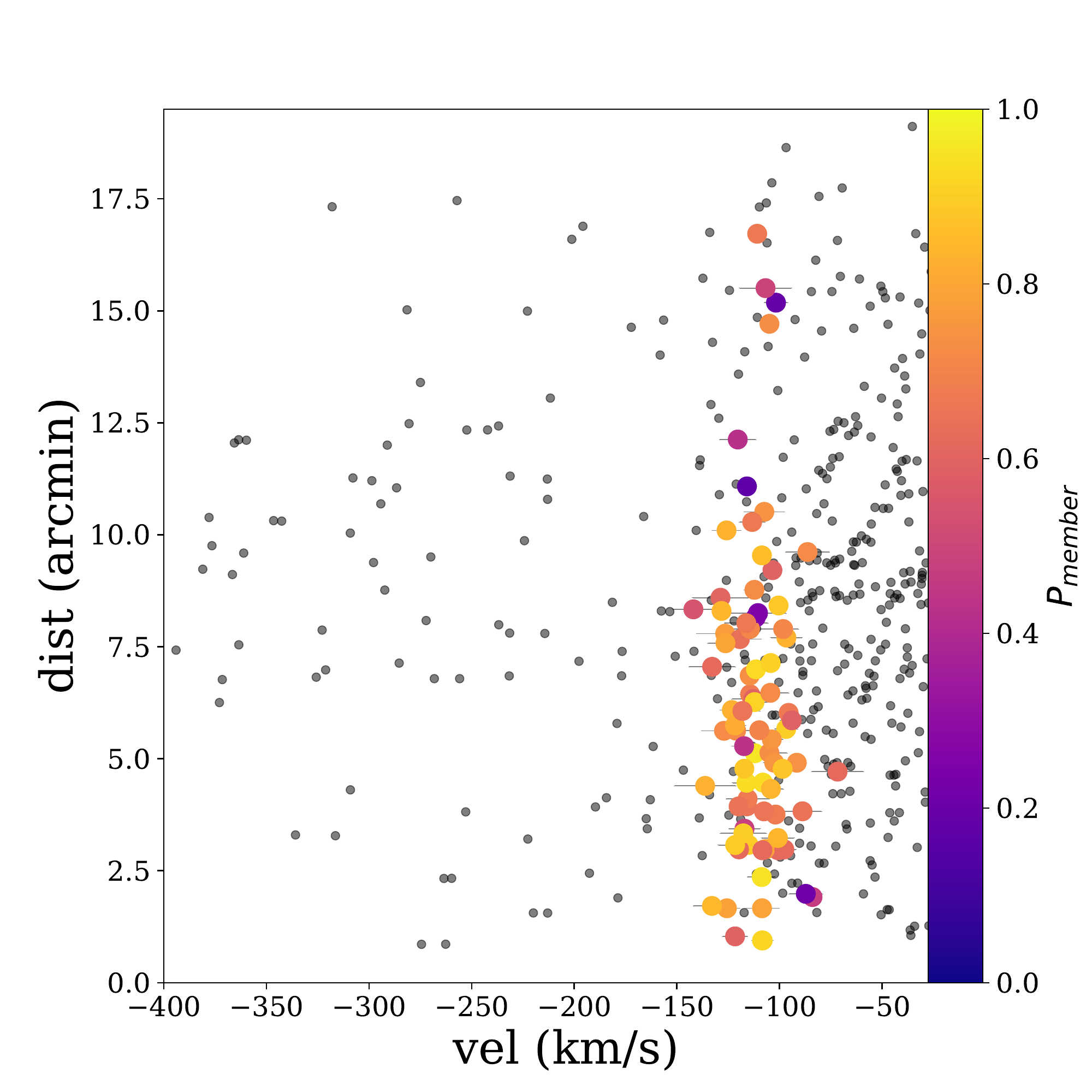}
     \includegraphics[angle=0,width=\columnwidth]{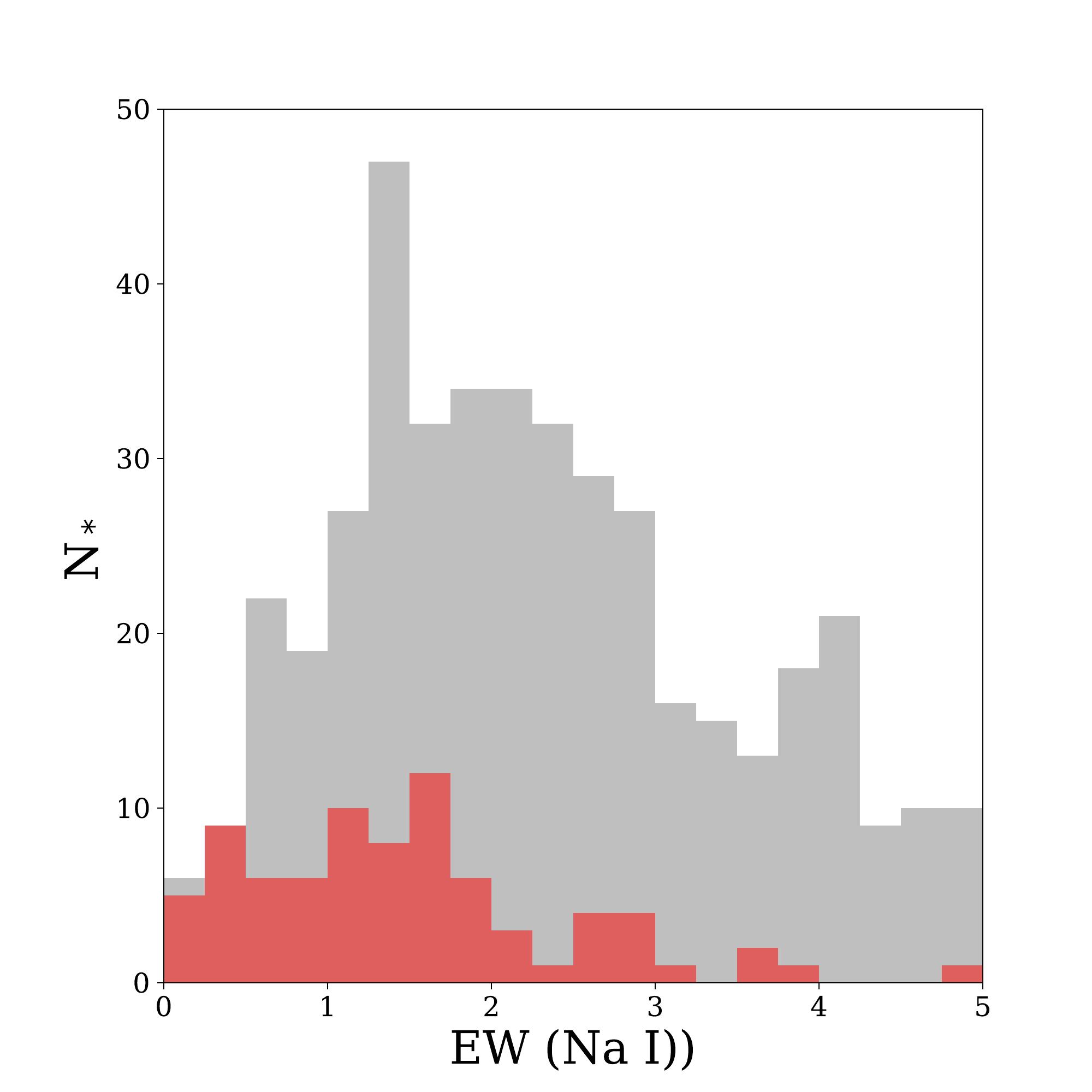}
    \caption{{\bf Top left:} Histogram of velocities for all spectroscopically observed stars (gray). Those most likely to be associated to And~XIX ($P_{\rm member}>0.1$) are highlighted in red, showing a cold distribution of stars at the systemic velocity of And~XIX. {\bf Top right:} Spatial distribution of all likely members of And~XIX, colour coded by velocity. The background grayscale shows the spatial distribution of RGB stars from Subaru imaging. {\bf Bottom left:} Velocity of all observed vs. distance from And~XIX (in arc minutes). {\bf Bottom right:}  Histogram showing the equivalent width of the sodium doublet (Na~I) for all stars observed with DEIMOS (grey histogram). Those with the highest probability of membership are shaded red.}
    \label{fig:kin}
\end{figure*}

\subsection{Membership determination}
\label{sec:members}

With a systemic velocity of $v_r\sim-110\kms$ \citep{collins13}, And~XIX member stars can be difficult to distinguish from MW foreground stars based on velocities alone (fig.~\ref{fig:kin}). To determine which stars within our DEIMOS sample are bonafide members, we employ a probabilistic approach following the procedures of \citet{collins13} and \citet{tollerud12}. We summarise this technique below. 

This method assigns the probability of membership of a given
star to the dwarf galaxy based for each of three criteria: (1) the star's
position on the color magnitude diagram of the dwarf galaxy, $P_{\mathrm {CMD}}$, (2) the distance of the star from the centre of the dwarf galaxy, $P_{\rm dist}$, and (3) the velocity of the star,
$P_{\rm vel}$. The probability of membership can then be expressed as a
multiplication of these three criteria:

\begin{equation}
P_{\mathrm {member}}\propto P_{\mathrm {CMD}}\times P_{\mathrm{ dist}}\times P_{\mathrm {vel}}
\end{equation}

$P_{\rm  CMD}$ is determined using the colour magnitude diagram (CMD) of And XIX. We
implement a method based on that of \citet{tollerud12}, using an isochrone to isolate those stars most likely to be associated with And~XIX. In fig.~\ref{fig:CMDs}, we show the CMD of And~XIX. We have overlaid an old, metal poor isochrone from the {\sc Parsec} stellar evolutionary models ([Fe/H]$=-1.8$, $[\alpha/$Fe]$=0.0$, age = 12~Gyr, shifted to a distance modulus of $m-M=24.57$ \citealt{bressan12,conn13}) that well represents the RGB of the dwarf galaxy. To assess the probability of a star being associated to And~XIX, we measure the minimum distance of a star from this isochrone ($d_{\rm min}$), and assign a probability using the following equation:

\begin{equation}
P_{\rm CMD}=\exp \left(\frac{- d_{\rm min}^2} {2  \sigma_{\rm CMD}^2}\right)
\label{eq:pcmd}
\end{equation}

\noindent where $\sigma_{\rm CMD}=0.1$. 

$P_{\rm dist}$ is determined using the known radial surface brightness profile of the dwarf, modelled as an exponential profile, using the half-light radius and ellipticity parameters for And~XIX as determined from PAndAS data \citep{martin16b}. $P_{\rm dist}$ can simply be written as:

\begin{equation}
P_{\rm dist}=  \exp(-r^2/2r_{\rm h}^2)
\end{equation} 

\noindent where $r_h$ is the elliptical half-light radius, and would be equal to $r_{\rm half}$ for a perfectly spherical system. We modify both $r$ and $r_{\rm half}$ based on a stars angular position with respect to the dwarf's major axis, $\theta=34^\circ$ \citep{martin16c}, such that:

\begin{equation}
r_{(\rm h)} = \frac{r_{\rm half}(1-\epsilon)}{1+\epsilon\cos\theta}
\end{equation}

$P_{\rm vel}$
is determined by simultaneously fitting the
velocities of all observed stars assuming that 3 dynamically distinct components are present: the MW
foreground contamination ($P_{\rm MW}$, modeled as a single Gaussian with systemic velocity $v_{\rm MW}$ and velocity dispersion of $\sigma_{v,\mathrm{MW}}$), the M31 halo
contamination ($P_{\rm M31}$, modeled as a single, broad Gaussian with systemic velocity $v_{\rm M31}$ and velocity dispersion of $\sigma_{v,\mathrm{M31}}$), and a further, single
Gaussian component to represent the substructure of interest (in this case,
And~XIX, $P_{\rm A19}$) with an arbitrary systemic velocity, $v_r$ and velocity dispersion
$\sigma_v$:

\begin{figure*}
     \includegraphics[angle=0,width=\columnwidth]{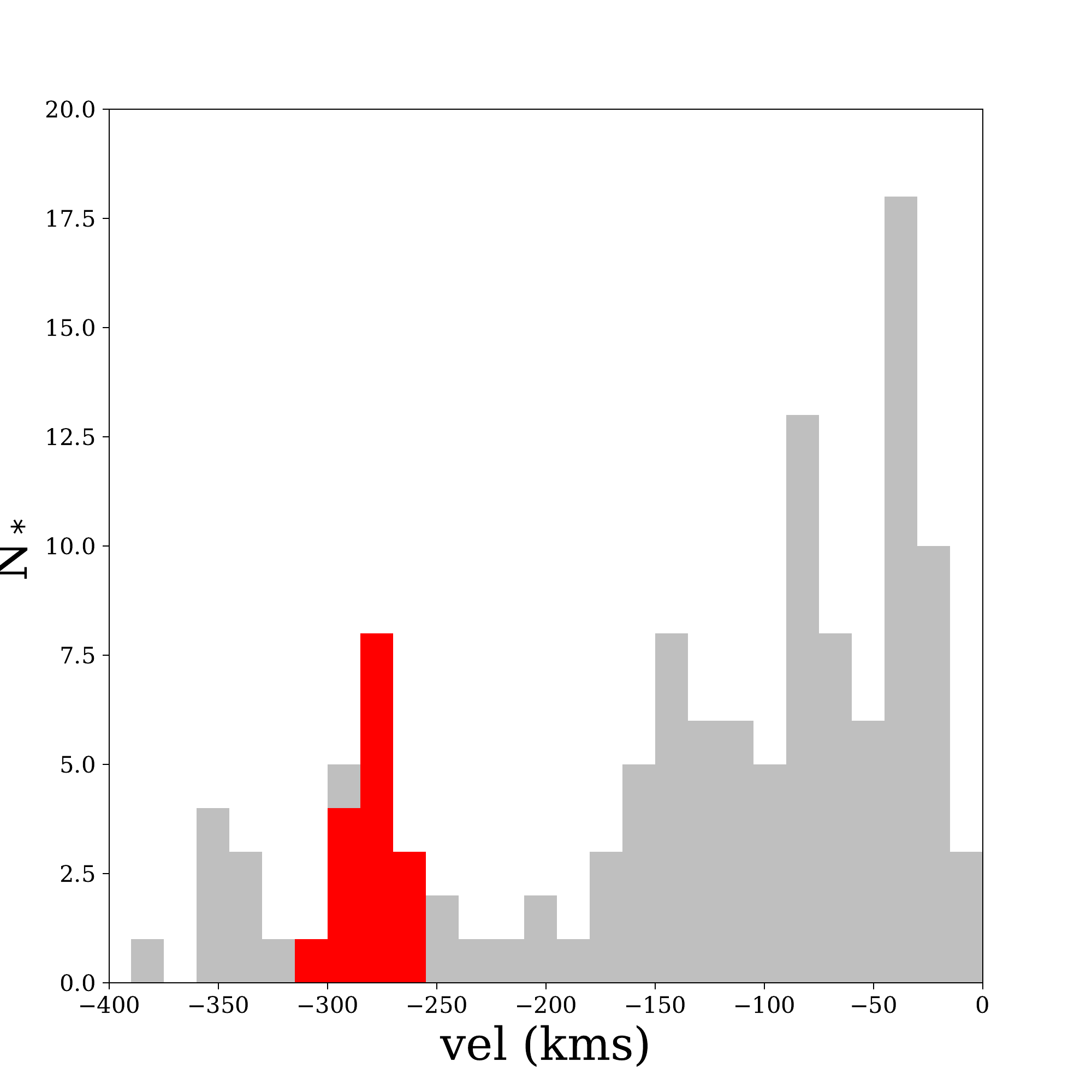}
     \includegraphics[angle=0,width=\columnwidth]{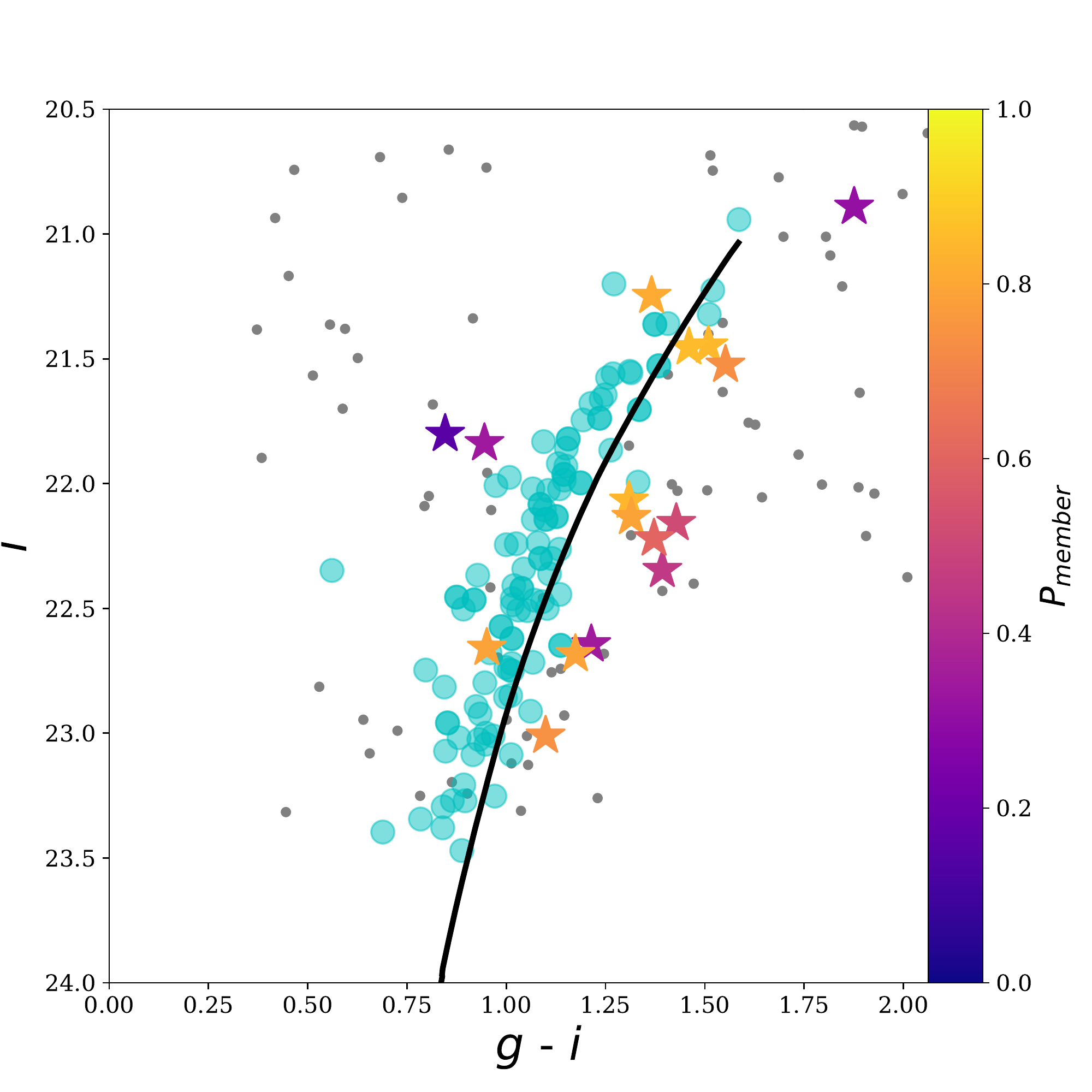}
    \caption{{\bf Left:} Histogram of velocities for all spectroscopically observed stars in our stream fields (gray). Those most likely to be associated with the stream($P_{\rm member}>0.1$) are highlighted in red. We see a cluster of 16 stars with $v\sim-280\kms$. {\bf Right:} PAndAS colour magnitude diagram for the stream region. Our likely members are shown as stars, and are colour coded by their probability. For reference, our probable members of And~XIX are also shown in this parameter space as cyan circles. We see that they appear more metal poor than the stream stars. A Dartmouth isochrone with an age of 12 Gyr, $[\alpha/{\rm Fe}]=+0.0$ and a metallicity of [Fe/H]$=-1.5$~dex is shown for reference.}
    \label{fig:stream}
\end{figure*}

\begin{equation} 
\begin{aligned}
  P_{\rm M31}=\frac{1}{\sqrt{2\pi(\sigma_{v,\mathrm{M31}}^2+\delta_{vr,i}^2)}}\times{\rm exp}\left[-\frac{1}{2}\left(\frac{v_{\mathrm{M31}}-v_{r,i}}{\sqrt{\sigma_{v,\mathrm{M31}}^2+\delta_{vr,i}^2}}\right)^2\right] \end{aligned},
\end{equation}

\begin{equation} 
\begin{aligned}
  P_{\rm MW}=\frac{1}{\sqrt{2\pi(\sigma_{v,\mathrm{MW}
      }^2+\delta_{vr,i}^2)}}\times{\rm exp}\left[-\frac{1}{2}\left(\frac{v_{\rm MW
        }-v_{r,i}}{\sqrt{\sigma_{v,\mathrm{MW}}^2+\delta_{vr,i}^2}}\right)^2\right] ,
\end{aligned}
\end{equation}

\begin{equation} 
\begin{aligned}
\label{eqn:pa19}
P_{\rm A19}=\frac{1}{\sqrt{2\pi(\sigma_{v}^2+\delta_{vr,i}^2)}}\times{\rm exp}\left[-\frac{1}{2}\left(\frac{v_{r}-v_{r,i}}{\sqrt{\sigma_{v}^2+\delta_{vr,i}^2}}\right)^2\right].
\end{aligned}
\end{equation}

A single Gaussian is an oversimplification for the Milky Way (see, e.g. \citealt{gilbert06,collins13}), but is adequate for this analysis, where we merely need to separate likely Milky Way stars from And~XIX.The likelihood function can then be simply written as:

\begin{multline}
\mathcal{L}_i(v_{r,i}, \delta_{vr,i}|\mathcal{P}) = (1 - \eta_{\rm MW} - \eta_{\rm M31}) \times P_{\rm A19}\\
+ \eta_{\rm MW} \times P_{\rm MW} + \eta_{\rm M31} \times P_{\rm M31}
\label{eqn:pdfsimple}
\end{multline}

\noindent where $\eta_{\rm MW}$ and $\eta_{\rm M31}$ are the fraction of our sample found within the Milky Way and M31 halo components of the model. We use the MCMC \texttt{emcee} code \citep{fm13a} to explore a broad parameter space for these components. We use uniform priors for each of our parameters (the velocities and dispersions for each population), constraining them to be within plausible physical ranges (see table~\ref{tab:priors} for details). In addition,  we set $0<\eta<1$ for all populations, with $\eta_{A19}+\eta_{MW}+\eta_{M31}=1$. In this kinematic-only analysis, we measure $v_{r,A19} =-108.8\pm1.3~\kms$ and $\sigma_{v,A19}=8.5^{+2.0}_{-1.8}~\kms$. Due to And~XIX's position in velocity space -- within the distribution of MW contaminants -- these values are likely to be close to the true values, but the velocity dispersion may be over-estimated. As such, this only a first step in determining the kinematic parameters. To measure the true values, we need to remeasure the dispersion using \texttt{emcee} with these probabilities as a weight to remove unwanted Milky Way contaminants (see \S\ref{sec:dynamics}).

\begin{table*}
	\centering
	\caption{Prior values used in our {\sc emcee} analysis}
	\label{tab:priors}
	\begin{tabular}{lcccc} 
			\hline
		Property & Prior  & & & \\
		& And~XIX & MW & M31 & Stream \\
		\hline
		$v_{r} (\kms)$ & $-140<v_{r}<-90$ & $-90< v_{r}<0$ & $-400<v_{r}<-200$  & $-300<v_r<-250$\\
		$\sigma_{v} (\kms)$ & $0<\sigma_{v}<50$ & $0<\sigma_{v}<150$ & $0<\sigma_{v}<500$ & $0<\sigma_{v}<50$ \\
		$\theta$ (deg) & $12<\theta<90$ & -- & --& --\\
		$\vgrad (\kms)$ & $-50 <\vgrad<50$& -- & -- & --\\
		\hline
	\end{tabular}
\end{table*}

Owing to the large on-sky size of And~XIX relative to the DEIMOS field of view, the radial probability, $P_{\rm dist}$, is a very weak indicator of membership probability for this dataset. However, the joint combination of CMD position and velocity turn out to be very powerful for eliminating MW contaminants from our sample. In fig.~\ref{fig:kin}, we show a velocity histogram of all our observed stars, and those with a probability of membership $P_{\rm member}>0.1$ are indicated in red. Now, the cold population of stars associated with And~XIX are easily seen.  In the right hand panel of fig.~\ref{fig:kin}, we also show the spatial distribution of our probable members, and we see they well trace the underlying imaging for And~XIX. In the appendix table~\ref{tab:allstars}, we present the probability of membership for every observed star, alongside its imaging and kinematic properties.

We implement one final check on the likely association of each star with And~XIX. Our spectra also cover the region of the sodium doublet (Na~I), located at $\sim8200$\AA. These lines are sensitive to the surface gravity of a star, and can be much stronger in dwarf stars than RGB stars (although there is significant overlap in the equivalent widths of lines at bluer colours, e.g. \citealt{gilbert06}). In the lower right panel of fig.~\ref{fig:kin} we show the measured equivalent width for all our DEIMOS sample (open histogram). Those that are most probable members based on their CMD position and velocities are shaded red. The vast majority of our stars have low equivalent widths ($EW_{\rm Na}<2$~\AA). Selecting on this property alone would remove a large swathe of contaminants, however it does not remove all of them, as we see that our non-members heavily populate this region of parameter space. We find that CMD position acts as a much stronger constraint on the membership probability of And~XIX stars, and removes  the majority of stars with strong Na I doublets. However, as one final quality cut, we excise all stars with a sodium equivalent width of $EW_{\rm Na}>2$~\AA\ (reducing the number of stars with $P_{\rm member}>0.1$ from 126 to 81).

For our stream fields, we use the same method, however we drop the distance probability, instead using only the systemic velocities and CMD position to determine likely members. We also drop our Na I cut, as none of the stream stars show significant absorption at the location of the Na I doublet. In fig.~\ref{fig:stream}, we see that the most likely members have a systemic velocity of $\sim-280\kms$. In the CMD, we see that they cluster redward of an isochrone of [Fe/H]$=-1.5$ (taken from the Dartmouth isochrones, \citealt{dotter08}). As such, they appear more metal rich than the likely And~XIX members (shown as cyan circles in the right hand panel of fig.~\ref{fig:stream}).

\subsection{The dynamics of the And~XIX system}
\label{sec:dynamics}

\begin{figure*}
     \includegraphics[angle=0,width=0.9\hsize]{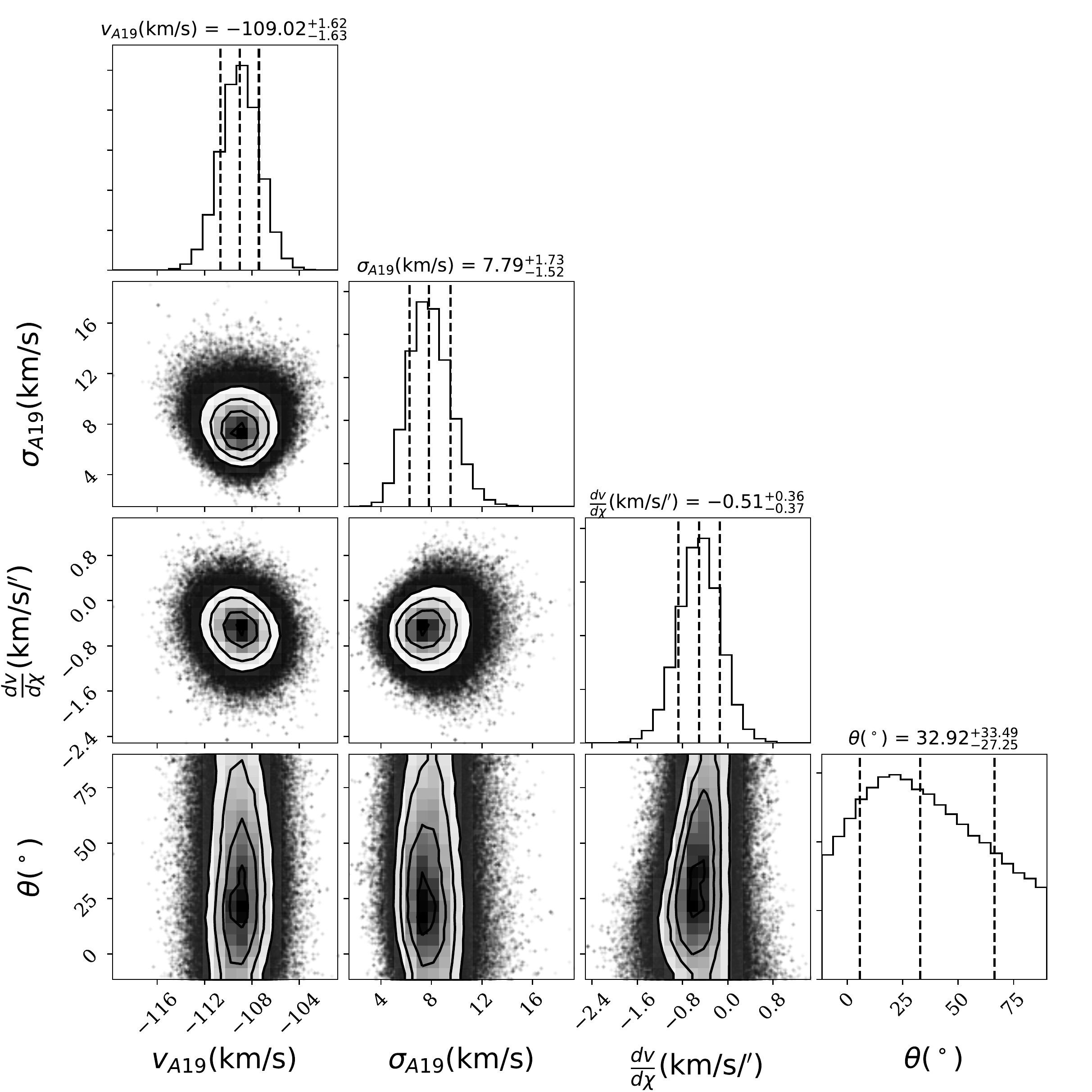}
    \caption{Two-dimensional and marginalized PDFs for the systemic velocity, velocity dispersion, velocity gradient, and position angle of this gradient for And~XIX. The dashed lines represent the mean value and 1$\sigma$ uncertainties.}
    \label{fig:mcmc}
\end{figure*}

Now that we have derived the probable membership of each And~XIX star, we can use their probabilities as weights in our analysis. We are interested in accurately determining And~XIX's systemic velocity ($v_r$), velocity dispersion ($\sigma_v$) and any velocity gradient that may be present in the system ($\vgrad$, where $\chi$ denotes the angular distance along the axis of rotation). For $\vgrad$, we follow the procedure of \citet{martin10} and \citet{collins17}. First, we define a likelihood function, $\mathcal{L}$, which describes a Gaussian population, with a velocity gradient that acts along a certain position angle, such that: 

\begin{multline}
\log~\mathcal{L}_{A19}\Bigg(v_{r,i}|\frac{{\rm d} v}{{\rm d}\chi},\langle v_r \rangle,\theta,\sqrt{\sigma_{vr}^2 + \delta_{vr,i}^2}\Bigg) = -\frac{1}{2}\sum_{i=0}^N  \log(\sigma^2) \\
+ \Bigg(\frac{\Delta v_{r,i}^2}{2\sigma^2}\Bigg) +  \log(2\pi) + \log(P_{{\rm member}, i})
\label{eqn:grad}
\end{multline}

\noindent where $\sigma=\sqrt{\sigma_{v}^2+\delta_{v,i}^2}$ is the combination of the underlying velocity dispersion of And~XIX ($\sigma_{v}$) and the velocity uncertainty of individual stars, $\delta_{v,i}$, and $P_{{\rm member}, i}$ is the probability of membership of the $i-$th star. Then, the velocity difference between the $i-$th star and a velocity gradient, $\vgrad$, acting along direction $y_i$ is defined as:

\begin{equation}
\Delta v_{r,i} = v_{r,i} - \vgrad y_i + \langle v_r \rangle
\end{equation}

\noindent The distance of a star from the centre of And~XIX in $X$ and $Y$ coordinates, centered on the dwarf, is $X_i = (\alpha_i-\alpha_0)~\mathrm{cos}(\delta_0)$, $Y_i =\delta_i-\delta_0$, where $\alpha_0, \delta_0$ ($\alpha_i, \delta_i$) is the position of the centre of And~XIX (the $i-$th star) in RA and dec. This is converted to an angular distance along an axis with a PA of $\theta$ such that $y_i = X_i~\mathrm{sin}(\theta) + Y_i~\mathrm{cos}(\theta)$. 

We then use {\sc emcee} to investigate the plausible parameter space for $v_r, \sigma_v, \vgrad$ and $\theta$.  In fig.~\ref{fig:mcmc} we show the results for And~XIX. We quote our final parameters as the median values from the {\sc emcee} posteriors, and the uncertainties are the $1\sigma$ confidence intervals from the posterior distribution. We find that \vrav, consistent with the value of $\langle v_r\rangle=-111.6^{+1.6}_{-1.4}~\kms$ from \citet{collins13}. We find a velocity dispersion from our updated sample of \sigav, which is higher than the value reported in \citet{collins13} of $\sigma_v=4.7^{+1.6}_{-1.4}~\kms$. Taking our uncertainties into account, these values are discrepant at the level of $1.5\sigma$. Given the small sample from which the previous kinematics were derived (24 stars), such a difference is not necessarily surprising. Further, when we examine the physical location of the original sample, their lower velocity dispersion is perhaps expected. We return to this in \S\ref{sec:mass}. 

Our analysis also favours a marginal velocity gradient in And~XIX of \vg ( equivalent to $\vgrad = -2.1\pm1.7~\kms$kpc$^{-1}$, using the distance to And~XIX of 821~kpc, \citealt{conn12b}), although this is detected with only slightly more than a 1$\sigma$ confidence. This gradient is essentially aligned with the major axis, with \thb\ (cf. the position angle of the galaxy of $\theta =34\pm5\deg$ \citealt{martin16c}), as one might expect for normal rotation.

Given the high fraction of Milky Way stars with similar velocities to And~XIX, there is a chance that our final kinematic measurements are affected by contaminants. To test this, we rerun our analysis with a very strict probability cut of $P_{\rm member}> 0.7$, which produces a sample of only 30 stars. We find no statistically significant difference in our results. The velocity dispersion lowers slightly to $\sigma_v=6.5\pm2.0~\kms$, well within the uncertainties of the above analysis. The velocity gradient increases to $\vgrad=-1.3\pm0.6~\kms$, but again this is consistent with our findings from the full sample of likely And~XIX members.

\begin{figure*}
     \includegraphics[angle=0,width=0.33\hsize]{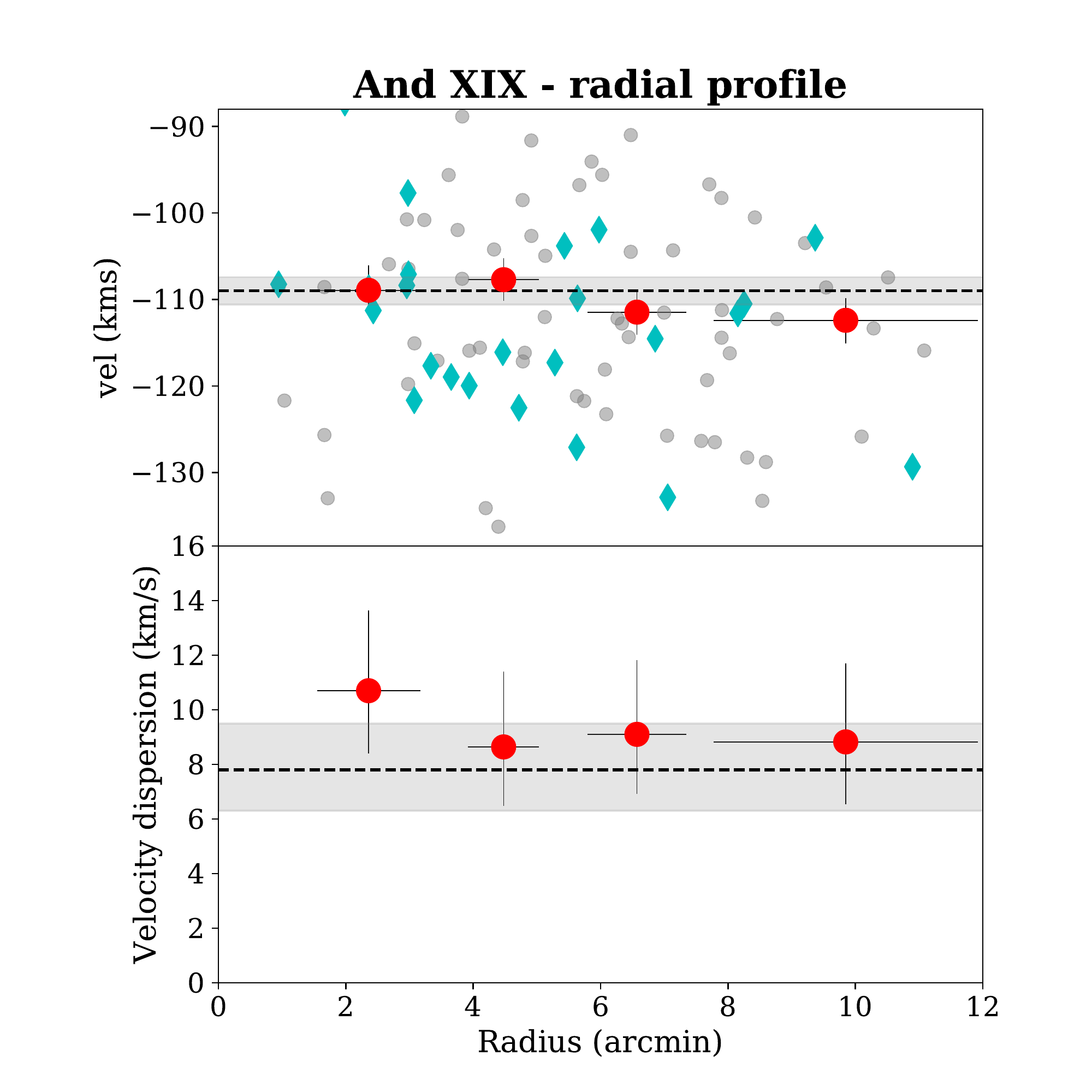}
     \includegraphics[angle=0,width=0.33\hsize]{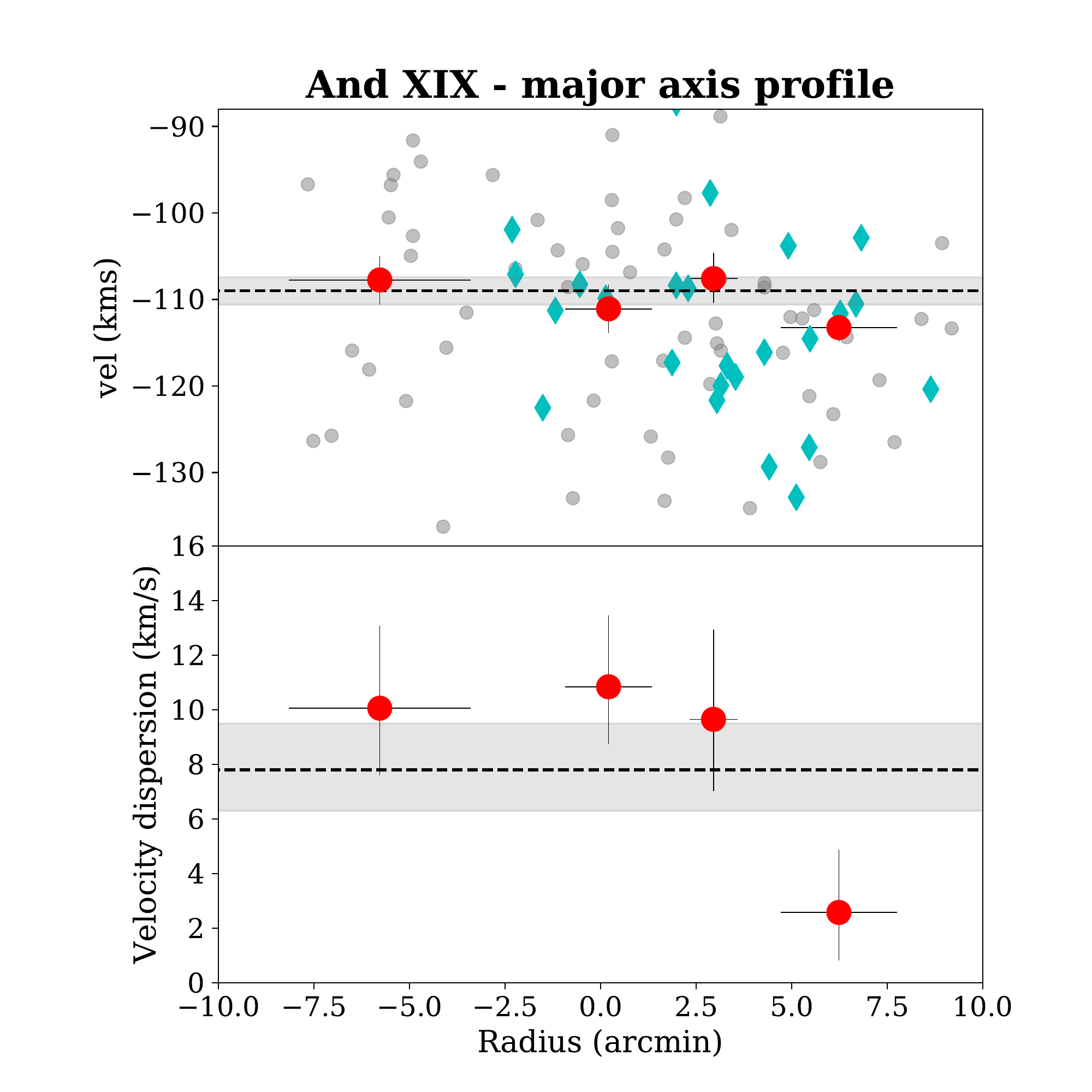}
     \includegraphics[angle=0,width=0.33\hsize]{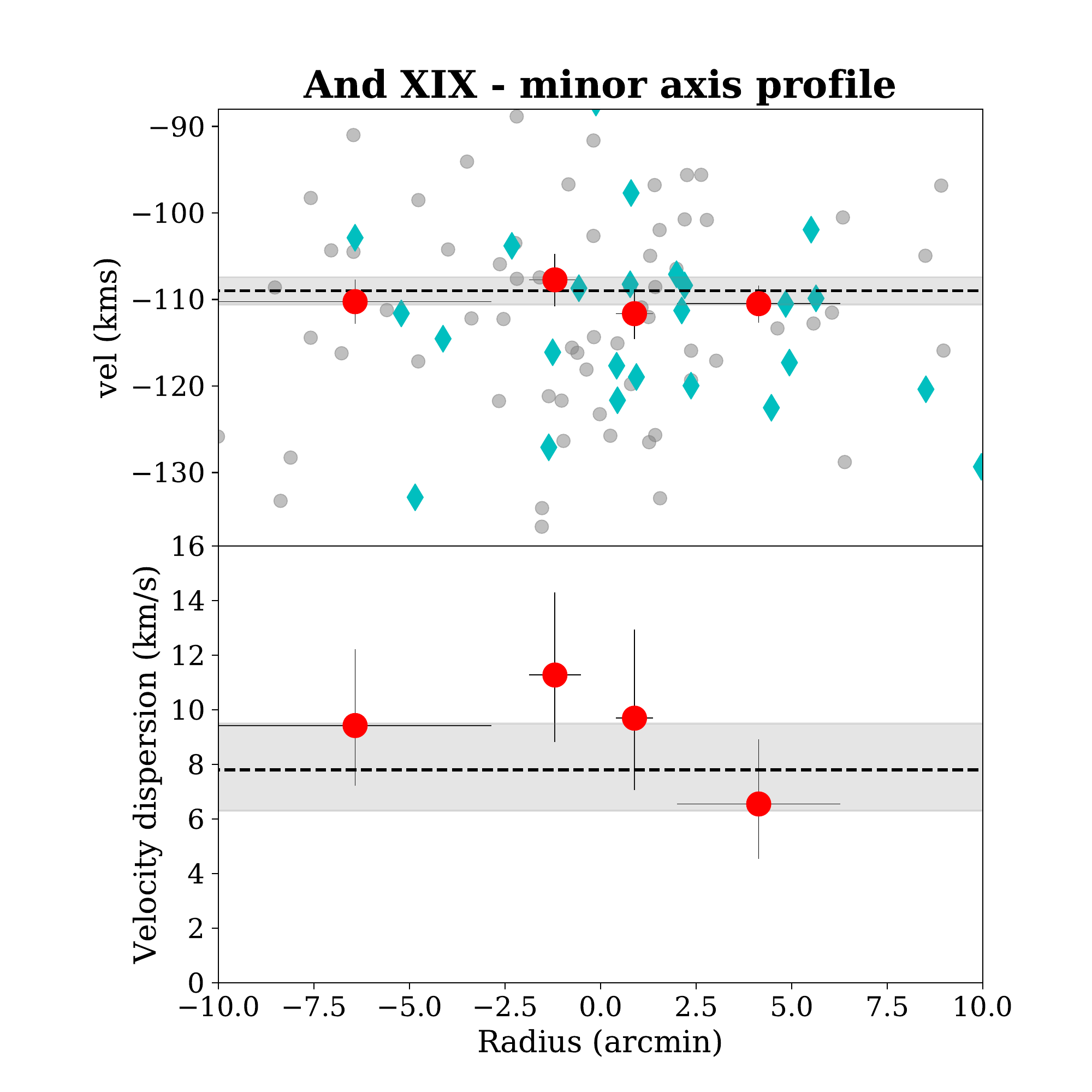}
    \caption{Kinematic profiles of And~XIX, showing how the systemic velocity and velocity dispersion of the galaxy behaves as a function of radius (left),  position along the major (middle) and minor axis (right). In each case, the dashed lines show the parameters derived for $v_r$ and $\sigma_v$ from our MCMC routine, with the shaded region showing the $1\sigma$ uncertainty, The radial and minor axis profile are largely flat, whereas the velocity gradient can be seen along the major axis profile. The outer most bin shows a lower systemic velocity, and well as a significantly lower velocity dispersion than the remainder of And~XIX. In each panel, we show the full sample of members as gray points. Those that were used in our initial study, \citet{collins13} are highlighted as blue diamonds.}
    \label{fig:prof}
\end{figure*}

We construct kinematic profiles of And~XIX as a function of distance from the centre, as well as along the major and minor axes (fig.~\ref{fig:prof}) to see how the systemic velocity and velocity dispersion vary across the galaxy. We do this by splitting our data into equal size bins that trace the radial, major and minor axes of the galaxy. We simplify eqn.~\ref{eqn:grad} by removing the velocity gradient term, and model each bin as a single Gaussian, as in eqn.~\ref{eqn:pa19}. We include only stars with $P_{\rm member}>0.1$ and $EW_{\rm Na}<2$\AA, resulting in $\sim20$ stars per bin. We run {\sc emcee} to determine the mean velocity and velocity dispersion within each bin. As expected, the radial profile of And XIX (top left, fig.~\ref{fig:prof}) is flat in terms of both the systemic velocity and dispersion, similar to MW dSphs (e.g. \citealt{walker07}). 

Along the major axis, we see the possible culprit for the marginal velocity gradient detected above. There is a slight decline in systemic velocity with increasing position along the major axis. Interestingly, the largest outlier in velocity from the average systemic (at $\sim+6^\prime$, or $+1.3$~kpc) is coincident with a significant drop in the velocity dispersion. This final bin has a velocity dispersion of only $\sigma_v=2.6^{+2.3}_{-1.8}~\kms$, an outlier at almost the $3\sigma$ level from the mean dispersion of And~XIX, and in tension with each of the other 3 bins at the $\gtrsim1.5\sigma$ level. Cold dips in dispersion profiles of dwarf galaxies have previously been linked to potential substructure. A similarly significant dip was noted in the radial profile of Andromeda II, and it was interpreted as a signature of a low mass merger in this system \citep{amorisco14a}.

Given And~XIX's inconvenient systemic velocity, which strongly overlaps with the Milky Way foreground, a thorough investigation of these offset stars is warranted before interpreting the significance of this feature. Are the stars in the cold `dip' they truly members of And~XIX? Or could they be contaminants? To address this, we inspect the position in the CMD for these low $v_r$ stars, and their individual spectra, to look for any signs that they are Milky Way foreground dwarf stars. We find that all these outliers have (1) a high probability of membership and (2) tightly cluster around the isochrone which best defines the And~XIX RGB. We also inspect the individual spectra for these low $\sigma_v$  stars. The Na~I doublet (located at $\sim8100$\AA) can be used to distinguish between foreground dwarf stars, and Andromeda giants (although somewhat imperfectly, as discussed above). Of the 24 probable And~XIX members in this radial bin, none show evidence for significant Na I lines (see fig.~\ref{fig:bin4}).  The lack of absorption at the location of the Na~I doublet in our probable And~XIX members, combined with their tight correlation with the And~XIX RGB leads us to conclude that these stars are not foreground dwarf contaminants. They could be Milky Way halo giants, in which case they would also not show any significant absorption. To test this hypothesis, we use the Besan\c{c}on stellar population model \citep{czekaj14} to simulate the Milky Way halo in the direction of And~XIX. We find that, over the entire area of our observations ($\sim0.22$~square degrees), we only expect to find 3 Milky Way halo giant stars in our full dataset of 627 stars with velocities and colours consistent with And~XIX. As such, we find that the stars in our analysis are likely associated with And~XIX itself. We will return to these stars in \S~\ref{sec:veldisc}.

\subsection{A revised  mass for And~XIX}
\label{sec:mass}

\begin{figure*}
     \includegraphics[angle=0,width=\columnwidth]{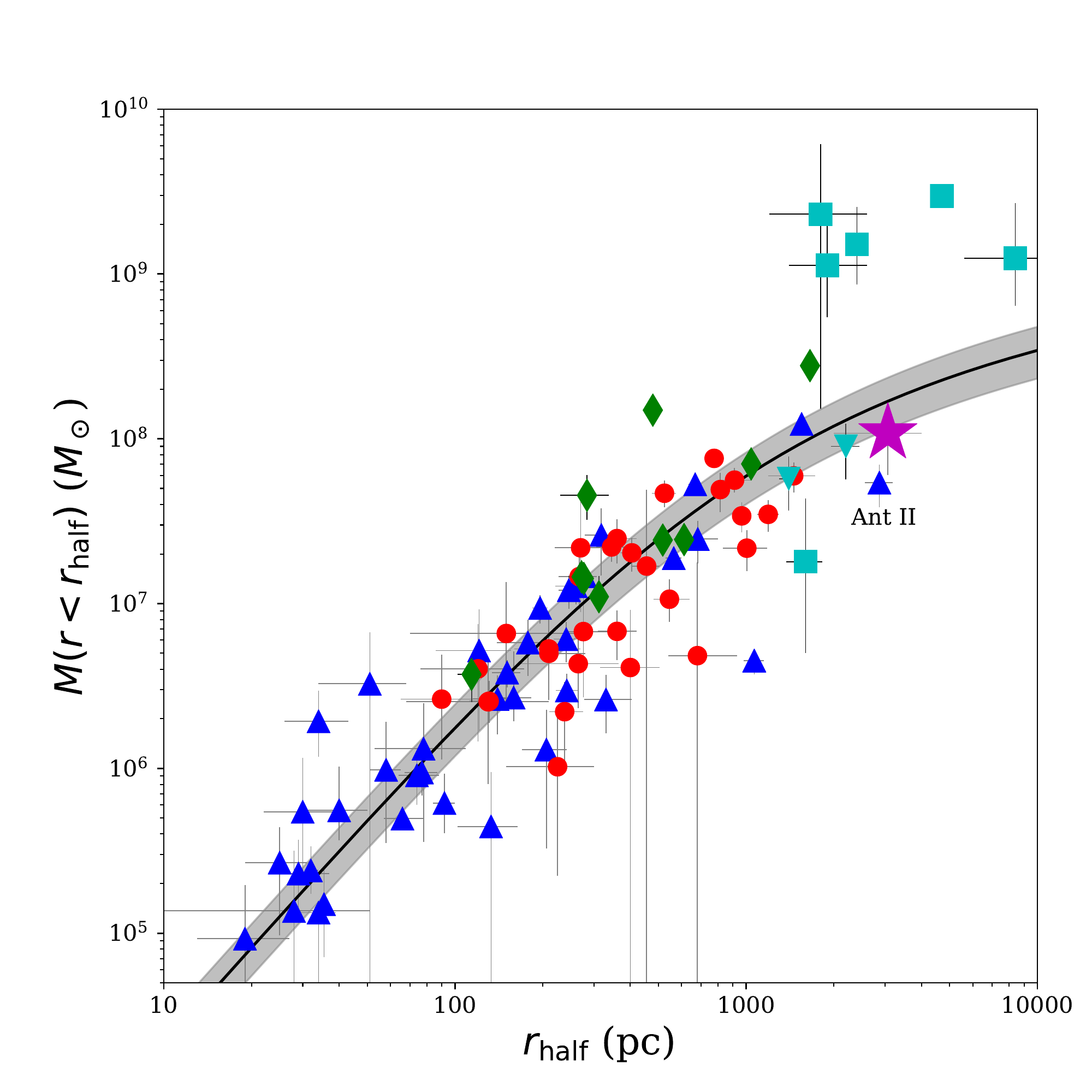}
     \includegraphics[angle=0,width=\columnwidth]{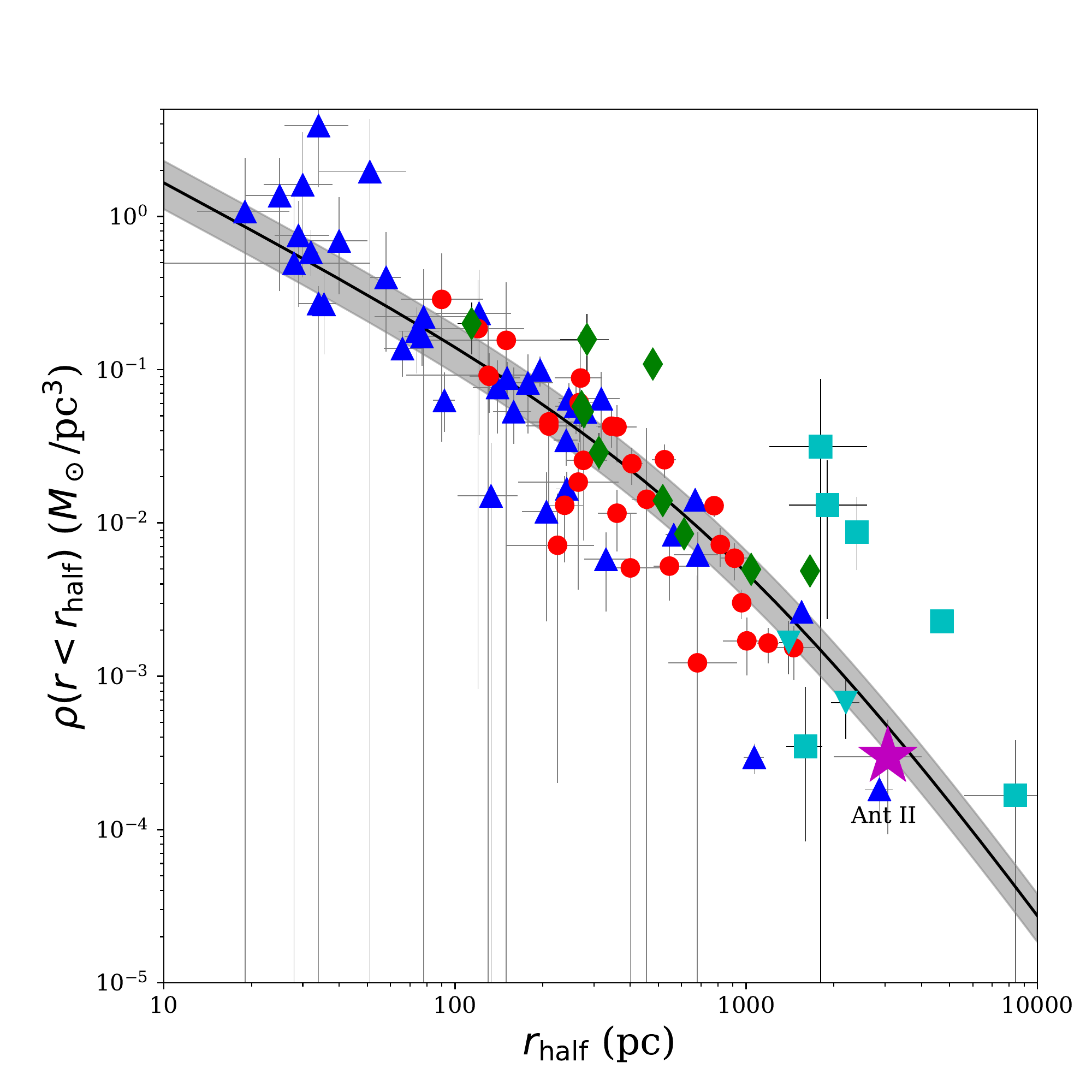}
    \caption{{\bf Left:} Mass within the half-light radius ($M(r<r_{\rm half})$ vs. $r_{\rm half}$ for Milky Way (blue triangles) and M31 (red circle) dSph galaxies, Local Group dwarf irregulars (green diamonds) and UDGs (cyan squares). The unusual UDG NGC 1052-DF2 is shown with two inverted triangles, see text for details. And~XIX is highlighted as a purple star. The black line represents the best fit NFW mass profile for Local Group dSphs from \citet{collins14}, with the grey shading indicating the $1\sigma$ scatter in this relation. And~XIX sits well within this regime, suggesting its halo is consistent with that of other dwarf galaxies. {\bf Right:} $r_{\rm half}$ vs. circular velocity as measured at the half light radius ($V_c$) for Local Group dwarf galaxies and UDGs. Circular velocity profiles from dark matter subhalos within the Aquarius simulations \citep{springel08} are overplotted. The kinematics of And~XIX place it in a low mass dwarf galaxy halo, with a maximum circular velocity of $\sim15\kms$.}
    \label{fig:mrh}
\end{figure*}

\begin{figure}
     \includegraphics[angle=0,width=\columnwidth]{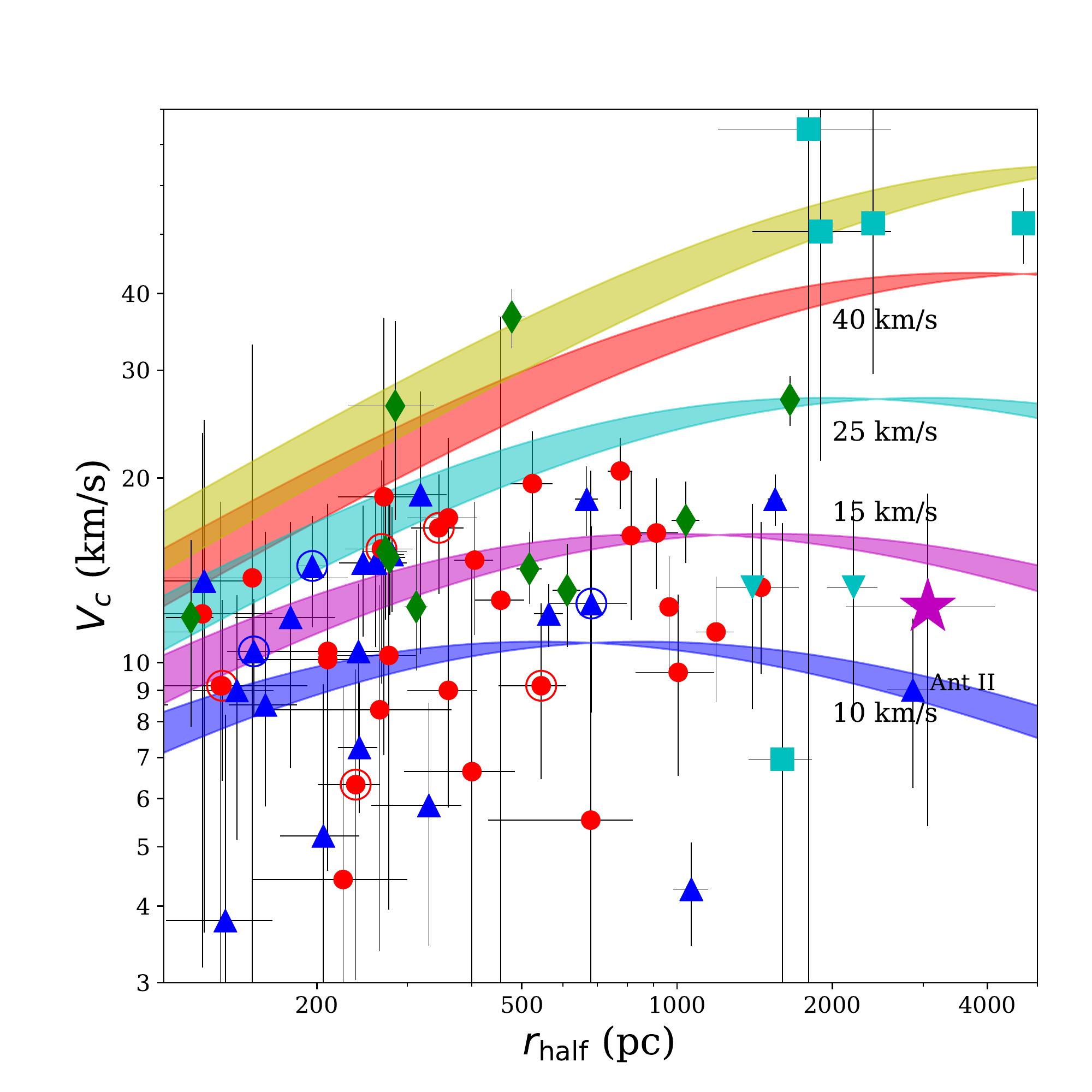}
    \caption{$r_{\rm half}$ vs. circular velocity as measured at the half light radius ($V_c$) for Local Group dwarf galaxies and UDGs. Circular velocity profiles from dark matter subhalos within the Aquarius simulations \citep{springel08} are overplotted. The kinematics of And~XIX place it in a low mass dwarf galaxy halo, with a maximum circular velocity of $\sim15\kms$.}
    \label{fig:vmax}
\end{figure}

Our initial analysis of the dynamics of And~XIX pointed to a stellar system embedded in a surprisingly low mass halo \citep{collins13,collins14}. Its velocity dispersion of $\sigma_v=4.7^{+1.6}_{-1.4}$ suggested it may inhabit a dark matter halo at or close to the molecular hydrogen cooling limit ($V_{\rm max}<10~\kms$, \citealt{koposov09}), at which point, galaxy formation is supposed to be very inefficient. This dispersion was measured from a sample of 24 members. Here, we measure velocities for  81 member stars, an increase of a factor of 3.4. From this new sample, we measure a higher velocity dispersion of \sigav, which is discrepant with our previous analysis at the level of $\sim1.5\sigma$. Measuring robust velocity dispersions from a small number of tracers can be very challenging (e.g. \citealt{martin18, laporte18}). This is especially true for And~XIX. With a systemic velocity of \vrav, this system lies within the broad kinematic distribution of the Milky Way foreground population, making a clean determination of membership difficult. By using a more robust methodology enabled by our larger sample size, we are better able to separate the kinematic profile of And~XIX from its contaminant population. Further, when we examine the physical positions of our original 2013 sample in fig.~\ref{fig:prof} (highlighted as blue diamonds), we note that these stars are located preferentially along one side of the major axis, coinciding with the low velocity dispersion bin. This would have led to a lower measured velocity dispersion. Indeed, when we re-run our MCMC analysis using only stars in the northern portion of And~XIX (i.e. with positive distance along the major axis), we recover a velocity dispersion of $\sigma_v=4.3^{+1.9}_{-2.0}~\kms$. This spatial coverage likely led to a biased measurement of the velocity dispersion, one that we have rectified with our larger sample.

We can use our new velocity dispersion to update the mass within the half-light radius of And~XIX, and determine if it is still an outlier in terms of its dark matter mass. The mass within the half-light radius has been shown to be a robust mass measure (e.g. \citealt{wolf10}) for dispersion supported systems, so long as they are in virial equilibrium. Given the dynamics of And~XIX discussed above, this may not be the case here and this could bias our mass inferences (we discuss this further in \S\ref{sec:disc}). However for completeness, we estimate the mass for And~XIX here, and compare it to other Local Group dwarf galaxies. We use the mass estimator of \citet{walker09a}, where:

\begin{equation}
M(r< r_{\rm half})=580 (M_\odot {\rm pc}^{-1}{\rm km}^{-2} s^2)~ r_{\rm half}\sigma_v^2 .
\end{equation}

\noindent We take $r_{\rm half}=3065^{+935}_{-1065}$~pc from \citet{martin16c}. This gives an enclosed mass of $M(r< r_{\rm half})=1.1\pm0.5\times10^8M_\odot$. In fig. ~\ref{fig:mrh}, we plot this as a function of half-light radius, and compare it with other Local Group dSphs (red circles, blue triangles, \citealt{walker09b, tollerud12, tollerud13, collins13, collins15,collins17, kirby15a, martin14b,martin16a,martin16b,caldwell17}), isolated dwarf galaxies (green diamonds, \citealt{mcconnachie12,kirby14,kirby17}) and UDGs (cyan squares,\citealt{vandokkum19b,toloba18,vandokkum19a,danieli19b}). The unusual UDG NGC1052-DF2 is shown using inverted triangles \citep{vandokkum18,danieli19a}. As its distance is disputed \citep{trujillo18,vandokkum18b}, we show the two possible positions for this galaxy in all our summary plots. In this parameter space, And~XIX looks fairly typical, falling just below the best fit mass relation for Local Group dSphs from \citet{collins14} (grey shaded region), suggesting it sits in a dwarf galaxy mass halo. We can also measure the central density of And~XIX's dark matter halo from this mass, and its half-light radius. As seen in the right hand panel of fig.~\ref{fig:mrh}, And~XIX has a very low average density, consistent with that of Ant~II (labelled).

We convert the mass within the half-light radius to a circular velocity, and compare this with rotation curves for dark matter halos from the Aquarius simulations (taken from \citealt{springel08}), shown in fig.~\ref{fig:vmax}. Despite its revised dispersion, And~XIX is still consistent with residing in a low mass halo, with $V_{\rm max}\sim15~\kms$. Such a low mass is consistent with similarly luminous (although more compact) dSphs in the Local Group (highlighted as encircled points in fig.~\ref{fig:vmax}). This could suggest that And~XIX is a ``puffed up'' dwarf galaxy, whose half-light radius has been increased as a result of tidal interactions, similar to what is discussed in \citet{carleton18}, \citet{amorisco19a} and \citet{torrealba18}.

\begin{figure}
     \includegraphics[angle=0,width=\columnwidth]{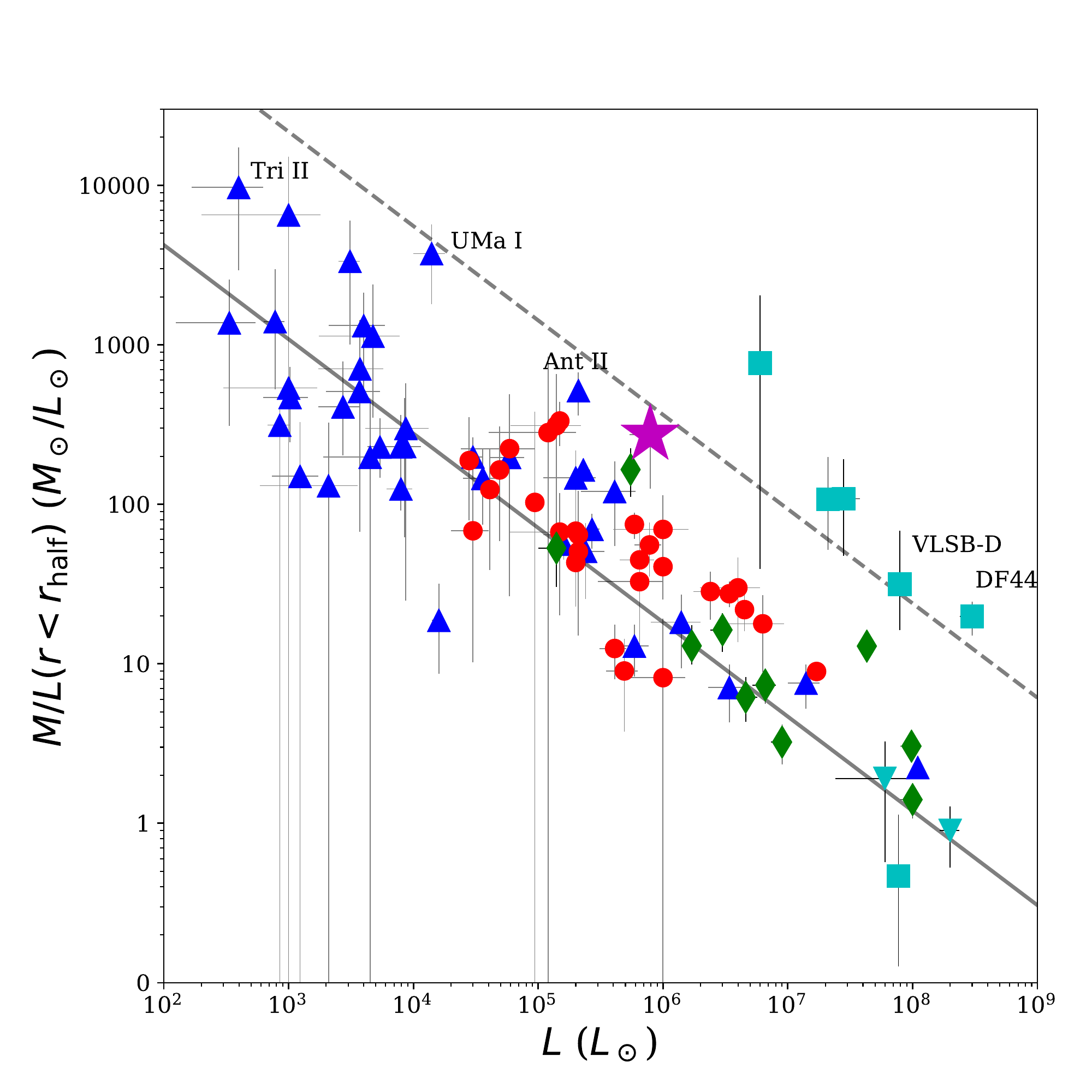}
    \caption{Luminosity vs. mass-to-light ratio within the half light radius ($M/L(r<r_{\rm half})$) for Local Group dwarfs and And~XIX (colours and symbols as in fig.~\ref{fig:mrh}). The solid gray line is a line of best fit to all the Local Group dwarf galaxies on the plot. The dashed line is simply this relation multiplied by 15. The UDGs, And~XIX, and several MW dSphs that are thought to be disrupting (labelled) are systematically offset from the main trend.}
    \label{fig:mll}
\end{figure}

 We measure the mass-to-light ratio of And~XIX within its half-light radius to determine how dark matter dominated it is. We calculate $[M/L]_{\rm half}= 278^{+146}_{-198} ~M_\odot/L_\odot$, implying that And~XIX is a dark matter dominated system. In fig.~\ref{fig:mll}, we plot $[M/L]_{\rm half}$ vs. $L$ for Local Group dwarf galaxies. Here, we see that the majority of systems follow a negative log-linear relationship in this parameter space, where lower luminosity galaxies have a higher mass-to-light ratio. A line of best fit, derived using a least-squares fitting procedure, to the Local Group population is shown as a solid gray line. And~XIX (magenta star) is an outlier to this relationship, appearing approximately $15\times$ more dark matter dominated (dashed line) than objects of a comparable stellar mass. Interestingly, a few other Local Group dSphs are also offset by this amount, most notably Ant~II, Ursa Major I (UMa I) and Triangulum II (Tri II). All these objects are suspected to be undergoing tidal stripping \citep{okamoto08,martin16b, torrealba18}. The UDG population is similarly offset to And~XIX too. This could suggest that And~XIX and some of the cluster UDG galaxies are undergoing tidal disruption or harassment, or that they are more dark matter dominated than more `typical' galaxies. We return to this in \S\ref{sec:udg}.

\begin{table}
	\centering
	\caption{The properties of And~XIX}
	\label{tab:a19prop}
	\begin{tabular}{ll} 
		\hline
		Property &  \\
		\hline
		$\alpha,\delta$ (J2000)$^1$ & 00:19:34.5, +35:02:01 \\
		$m_{V,0}$$^a$ & $14.5\pm0.3$ \\
		$M_{V, 0}$ $^a$& $-10.0^{+0.8}_{-0.4}$ \\
		Distance (kpc) $^b$& $821^{+32}_{-108}$\\
		$r_{\rm half}$ (arcmin) $^a$& $14.2^{+3.4}_{-1.9}$ \\
		$r_{\rm half}$ (pc)$^a$ & $3065^{+935}_{-1065}$\\
		$\mu_0$ (mag per sq. arcsec)$^a$ & $29.3\pm0.4$\\
		$L$ ($L_\odot$)$^a$ & $7.9^{+2.1}_{-3.9}\times10^5$ \\
		$v_r (\kms)$$^c$ & $-109.0\pm1.6$~kms$^{-1}$\\
		$\sigma_v (\kms)$ $^c$&  $7.8^{+1.7}_{-1.5}$~kms$^{-1}$\\
		$\vgrad (\kms/$arcmin)$^c$ & $-0.5\pm0.4$~kms$^{-1}$\\
		$\vgrad (\kms/$kpc) $^c$&  $-2.1\pm1.7$\\
		$M(r<r_{\rm half} (M_\odot)^c $& $1.1\pm0.5\times10^8$\\
		$[M/L]_{\rm half} (M_\odot/L_\odot)^c$ & $278^{+146}_{-198}$ \\
		$[{\rm Fe/H}]$ (dex) $^c$& $-2.07\pm0.02$ \\
		\hline
	\end{tabular}
	\\$^a$ \citet{martin16c},
	$^b$ \citet{conn12b},
	$^c$ This work
\end{table}

\subsection{The metallicity of And~XIX}
\label{sec:feh}

The low $S/N$ ratios for our spectra (ranging from $\sim3-15$\AA$^{-1}$ make detailed abundance calculations for the members of And~XIX difficult. The individual iron (Fe) lines are typically too weak to measure directly. However, there exists a well known empirical relationship between the strength of the Ca~II triplet absorption features and iron abundance ([Fe/H]) in RGB stars (e.g. \citealt{armandroff91}). In this work, we use the metallicity estimator from \citet{starkenburg10} to convert the equivalent widths of the Ca~II lines in our spectra into [Fe/H].

\begin{figure*}
     \includegraphics[angle=0,width=\columnwidth]{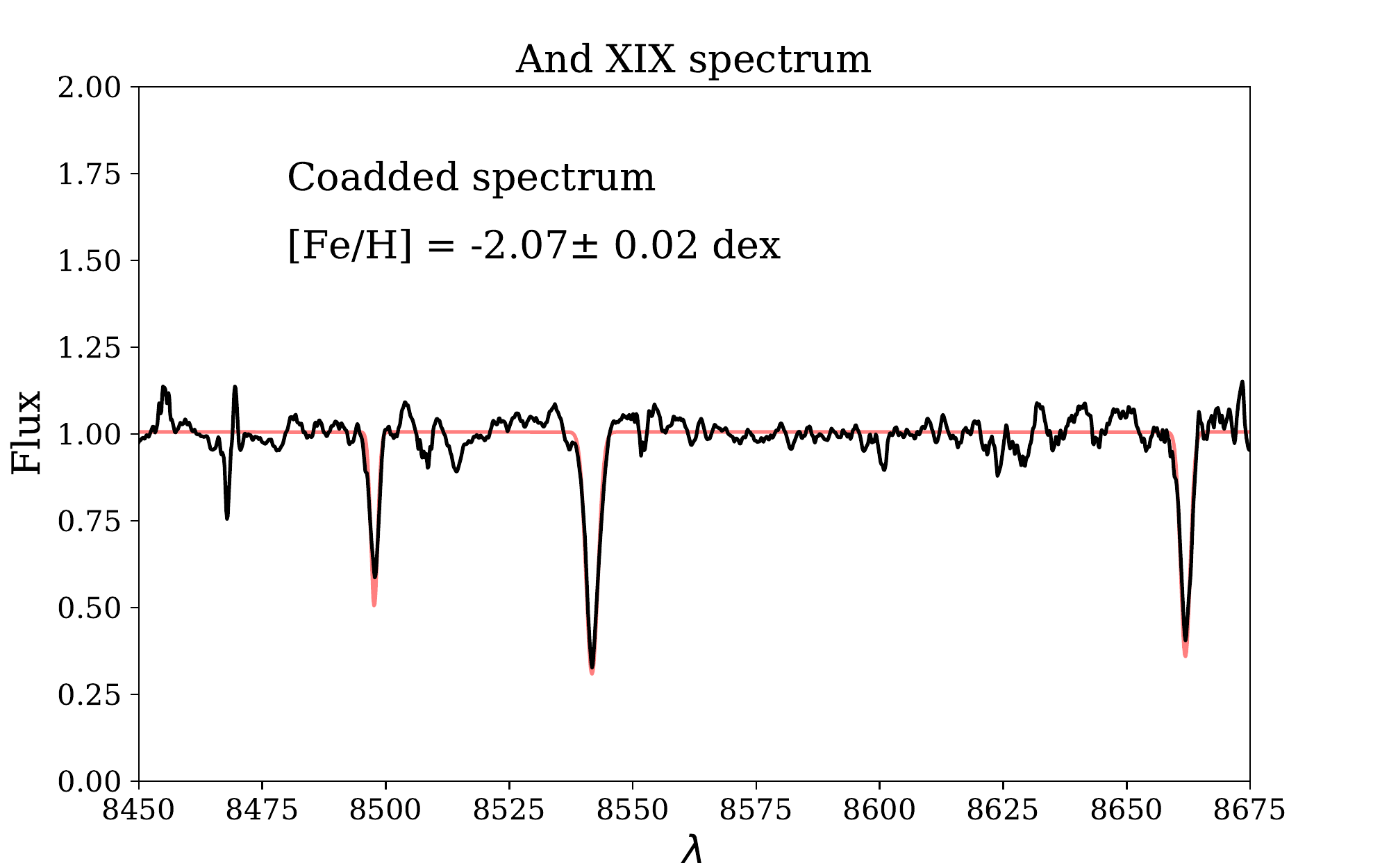}
     \includegraphics[angle=0,width=\columnwidth]{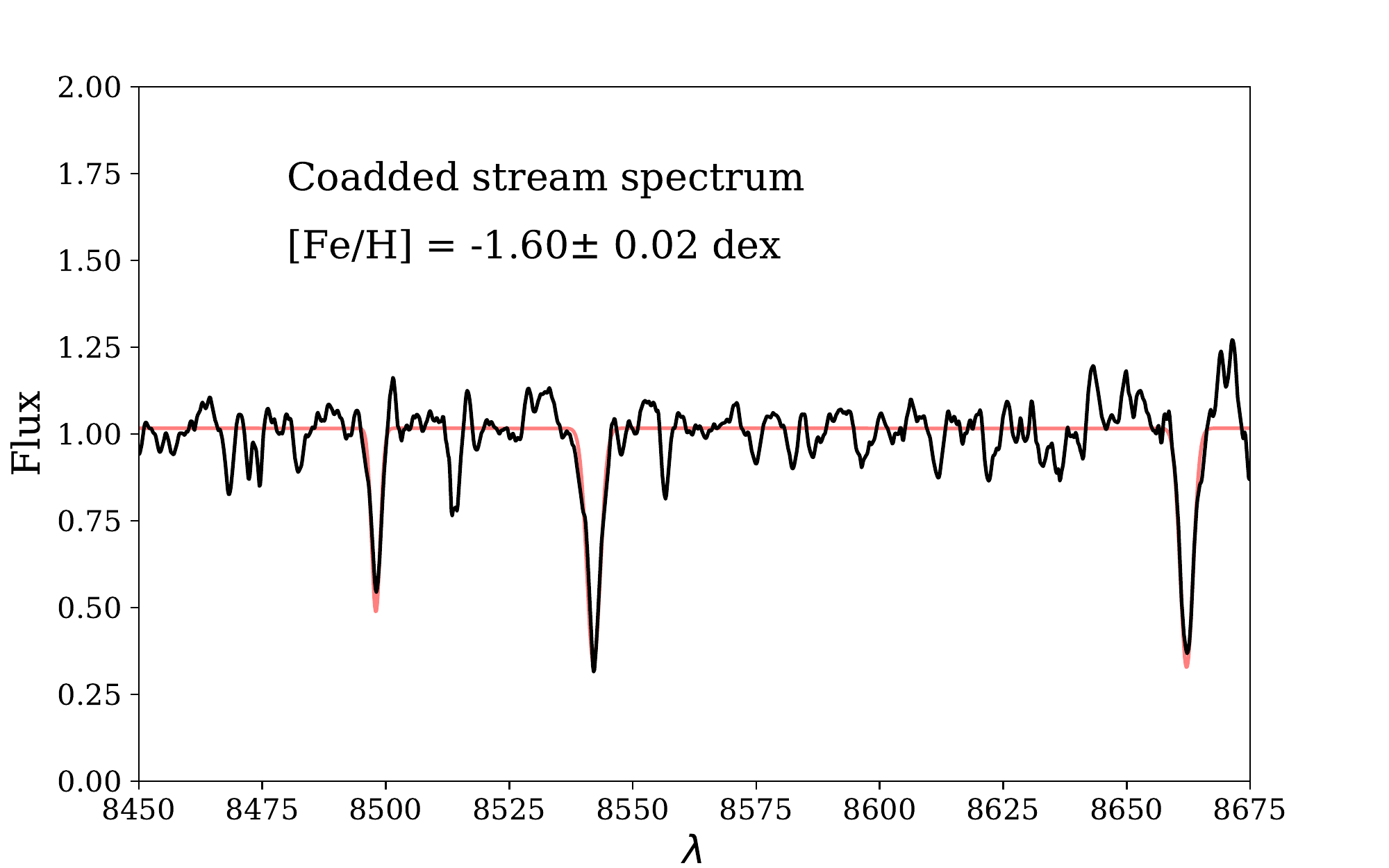}
    \caption{{\bf Left:} A co-added spectrum for all probable And XIX member stars. The red line shows our best fit to the continuum and Ca II lines. And XIX appears to be metal poor, with [Fe/H]$=-2.07\pm0.02$. {\bf Right:} The same, but for our likely stream members. We see that the stream appears more metal rich than And~XIX, with [Fe/H]$=-1.61\pm0.02$~dex, consistent with the imaging shown in fig.~\ref{fig:stream}.}
    \label{fig:coadd}
\end{figure*}

\begin{figure}
	\includegraphics[width=\columnwidth]{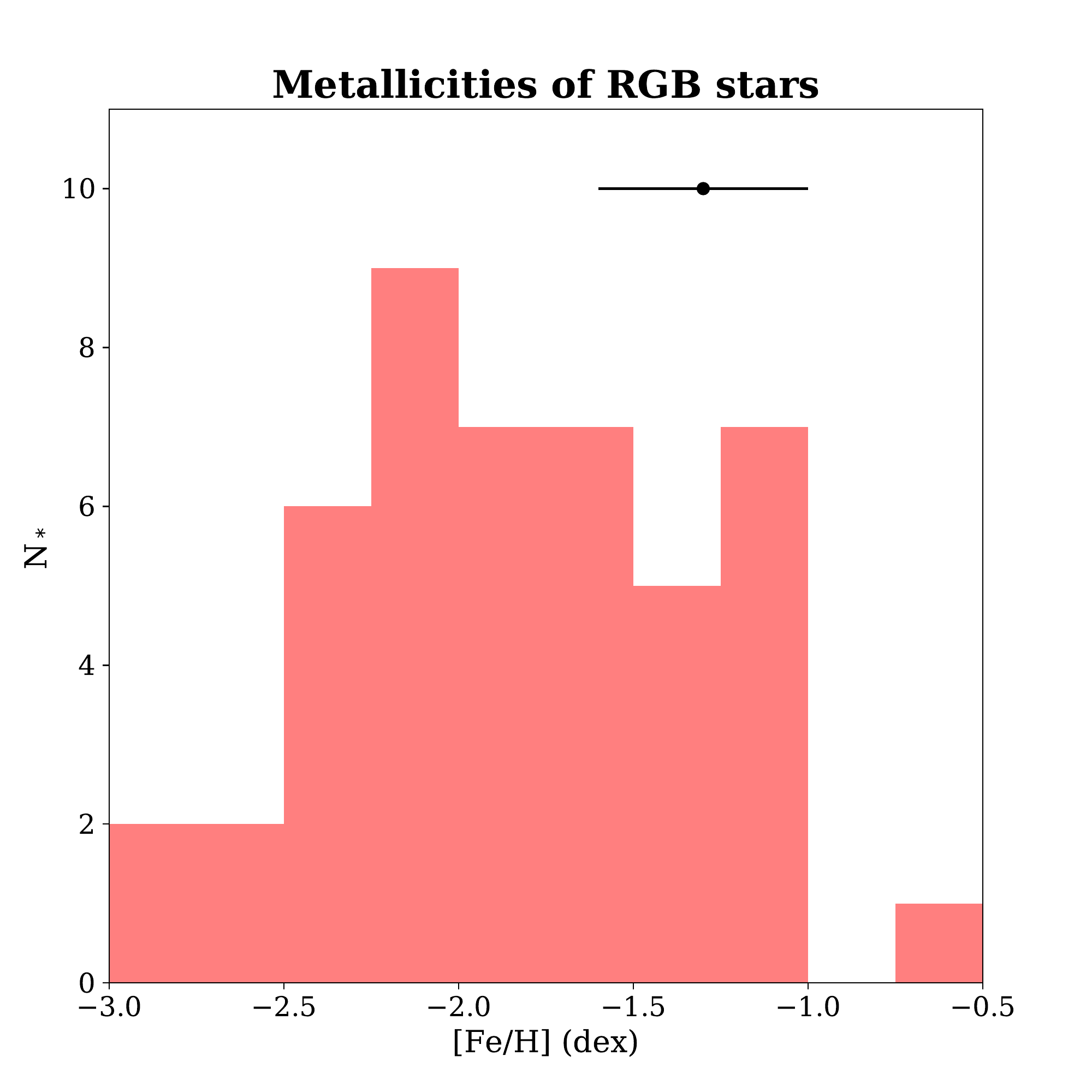}
    \caption{Metallicity distribution function for all likely ($P_{\rm member} >0.1$) members of And~XIX. Only stars with $S/N>5$\AA$^{-1}$ are included. The MDF is quite broad, however much of the spread is caused by the intrinsic uncertainties in the measurements. The mean metallicity peaks at [Fe/H]$\sim-2$~dex. Once accounting for measurement uncertainties (typical uncertainty shown as error-bar in plot), the spread in metallicities is $\sim0.5$~dex. }
    \label{fig:mdf}
\end{figure}

We present the individual metallicities for all stars with $S/N>5$\AA$^{-1}$ in table~\ref{tab:allstars}, and we show a histogram of the metallicities for stars with $P_{\rm member}>0.1$ in fig.~\ref{fig:mdf}. The MDF of And~XIX shows that the object is metal poor, with mean [Fe/H]$= -1.8\pm0.1$~dex. As our spectra are of low $S/N$, the uncertainties on individual metallicities are large ($>0.5$~dex). The spread in metallicity (once the artificial spread from the uncertainties is accounted for) is measured using the rms scatter, and is $\sigma_{\rm[Fe/H]}=0.5$~dex. This is consistent with metallicity spreads seen in similarly luminous galaxies \citep{kirby11}. The broad MDF in And~XIX is indicative of self-enrichment through extended star formation, similar to M31 dwarf galaxies where detailed star formation histories have been derived from deep HST imaging \citep{weisz14a,skillman17}.

Given the large uncertainties in metallicities from individual stars, we co-add the spectra of all probable members ($P_{\rm member}>0.1$) of And~XIX to derive a more accurate mean [Fe/H]. To construct the co-added spectrum, each star is corrected to its rest frame, weighted by its signal to noise, and then the weighted fluxes are summed. The mean [Fe/H] is then calculated using the same method as above. We measure And~XIX to be metal poor, with [Fe/H]$=-2.07\pm0.02$~dex, consistent with the mean from the MDF. If we employ a more stringent cut in probability ($P_{\rm member}>0.7$), we get a consistent (though slightly more metal poor) result of [Fe/H]$=-2.12\pm0.05$~dex. Given the small sample of stars with reasonable $S/N$, the uncertainty on the co-added spectrum is likely larger than implied by the fit uncertainties alone.

\begin{figure}
     \includegraphics[angle=0,width=\columnwidth]{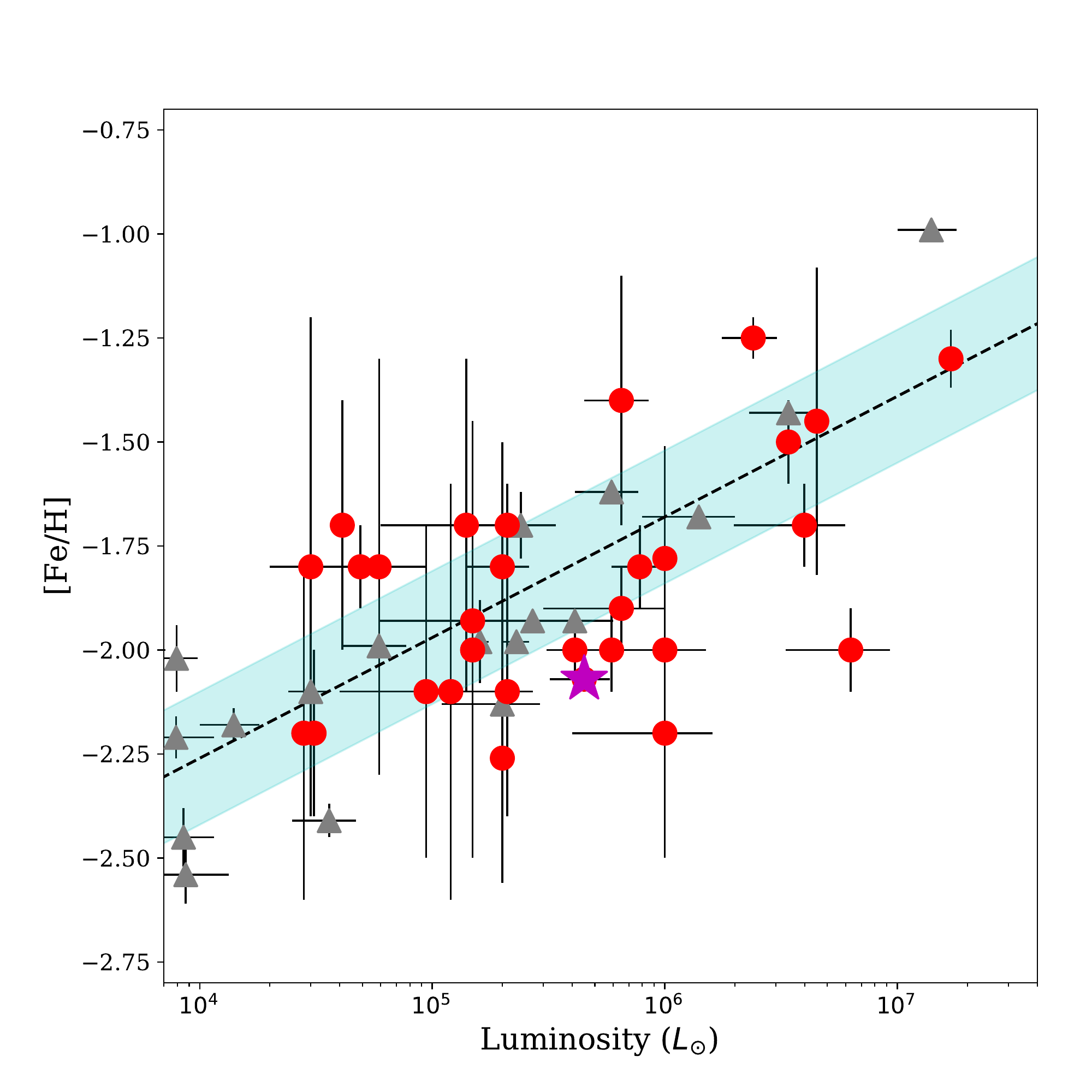}
    \caption{Here we show the luminosity vs. [Fe/H] relation for MW dSphs. Grey triangles are individual data, while the dashed line shows the \citet{kirby13a} relation for these systems. The scatter in the relation is shown as the cyan band. M31 dSphs with measured spectroscopic metallicities \citep{collins13,martin14b, ho15} are shown as red points. Our value for And~XIX based on the coadded spectrum is shown as a magenta star. }
    \label{fig:lfeh}
\end{figure}

In fig.~\ref{fig:lfeh}, we show the luminosity-metallicity relation for Local Group dwarf galaxies \citep{kirby13a}, where the dashed line represents the mean, and the cyan shaded region is the $1\sigma$ scatter around the relation. Our co-added result for And XIX is shown as the purple star. It seems that And XIX is more metal poor than would be expected for a galaxy of its luminosity, as it is an outlier to the Kirby et al. relation. We note that the MW dSphs typically have their metallicities measured directly from iron lines, where we have used the Ca II lines. This could lead to systematic offsets, however, we note that And~II, with a luminosity of $L=2.4\times10^6L_\odot$ and [Fe/H]$=-1.25\pm0.05$~dex \citep{ho15} is an outlier in the opposite direction to And~XIX, and its metallicity is also measured using the Ca~II lines. This low metallicity could imply that And~XIX has had a different star formation history when compared to Milky Way satellites of a comparable luminosity. Deep imaging of And~XIX with HST program 15302 (PI Collins) were taken in October 2018 in order to measure its star formation history, and these should help address the nature and evolution of And~XIX.

\subsection{An analysis of the stream feature}
\label{sec:stream}

To assess whether the stream feature seen to the west of And~XIX is truly associated, we analyse the kinematics and chemistry of the 16 probable members identified in fig.~\ref{fig:stream}. As already mentioned, the location of these stars in the PAndAS CMD implies this stream may be more metal rich than And~XIX, which would disfavour an association. While metallicity gradients in dwarf galaxies are not uncommon, typically the metallicity decreases with radius, implying that a stream would be either the same metallicity or lower than its progenitor.

\begin{figure}
     \includegraphics[angle=0,width=0.9\hsize]{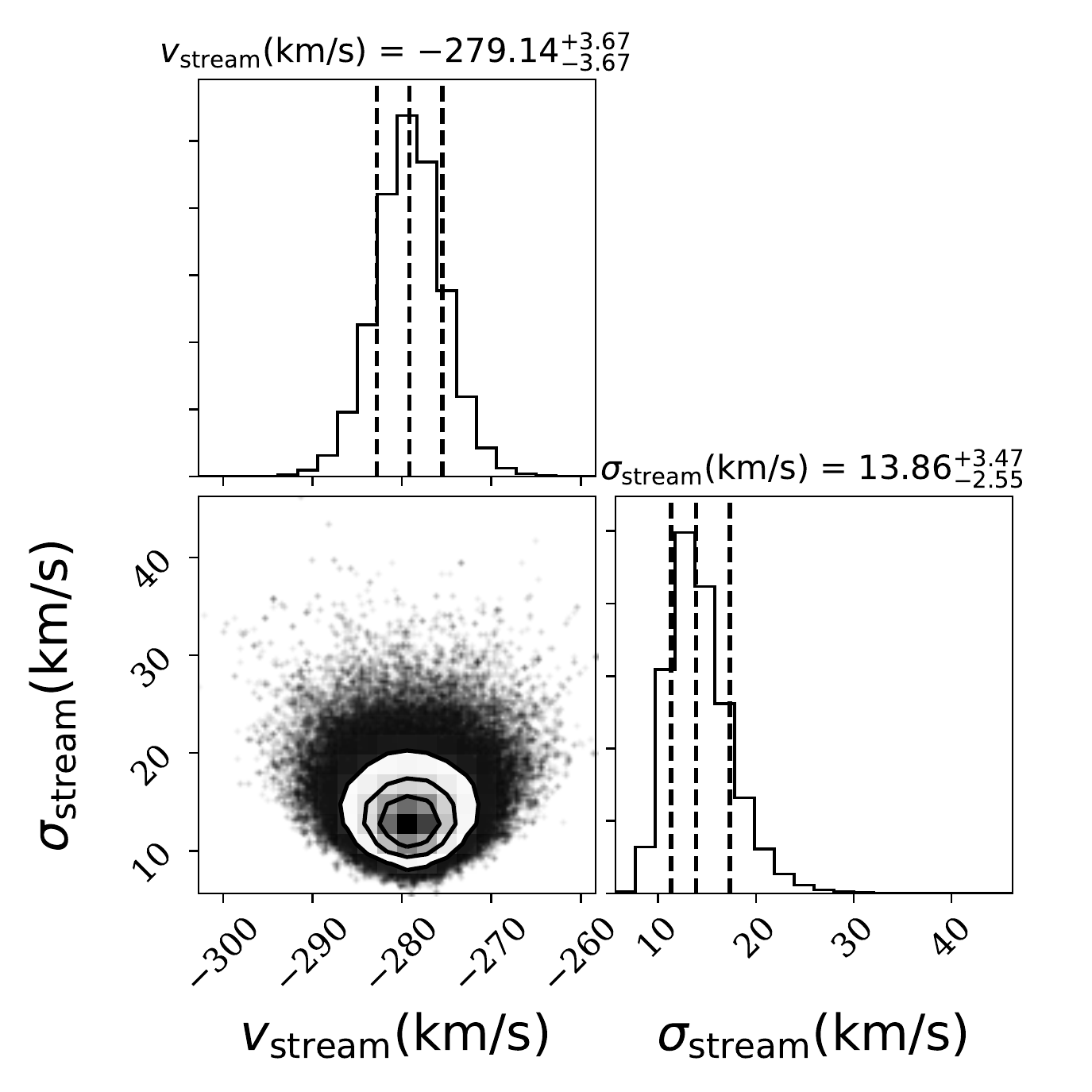}
    \caption{Two-dimensional and marginalized PDFs for the systemic velocity and velocity dispersion for the stream feature to the west of And~XIX. The dashed lines represent the mean value and 1$\sigma$ uncertainties.}
    \label{fig:smcmc}
\end{figure}

First, we determine the systemic velocity and dispersion of the stream using {\sc emcee}, as above (though neglecting a velocity gradient). Our results are shown in fig.~\ref{fig:smcmc}. We find $\langle v_r\rangle=-279.2\pm3.7~\kms$ and $\sigma_v=13.8^{+3.5}_{-2.6}~\kms$ for the stream. The velocity is offset from And~XIX by $\sim170~\kms$, and the dispersion is higher than that for And~XIX at a confidence of about $2\sigma$. Given the large separation on the sky of $\sim1-2$ degrees ($\sim15-30$~kpc), such a large velocity offset does not preclude the association of these two structures. However, when we co-add the spectra for our likely members (where we use a probability cut of $P_{\rm member}>0.3$), we find the stream to be significantly more metal rich than And~XIX, with [Fe/H]$=-1.61\pm0.02$~dex (fig.~\ref{fig:coadd}), just as we see in the imaging. As such, we find it highly unlikely that this stream is a result of the disruption of And~XIX.

\begin{figure}
     \includegraphics[angle=0,width=0.9\hsize]{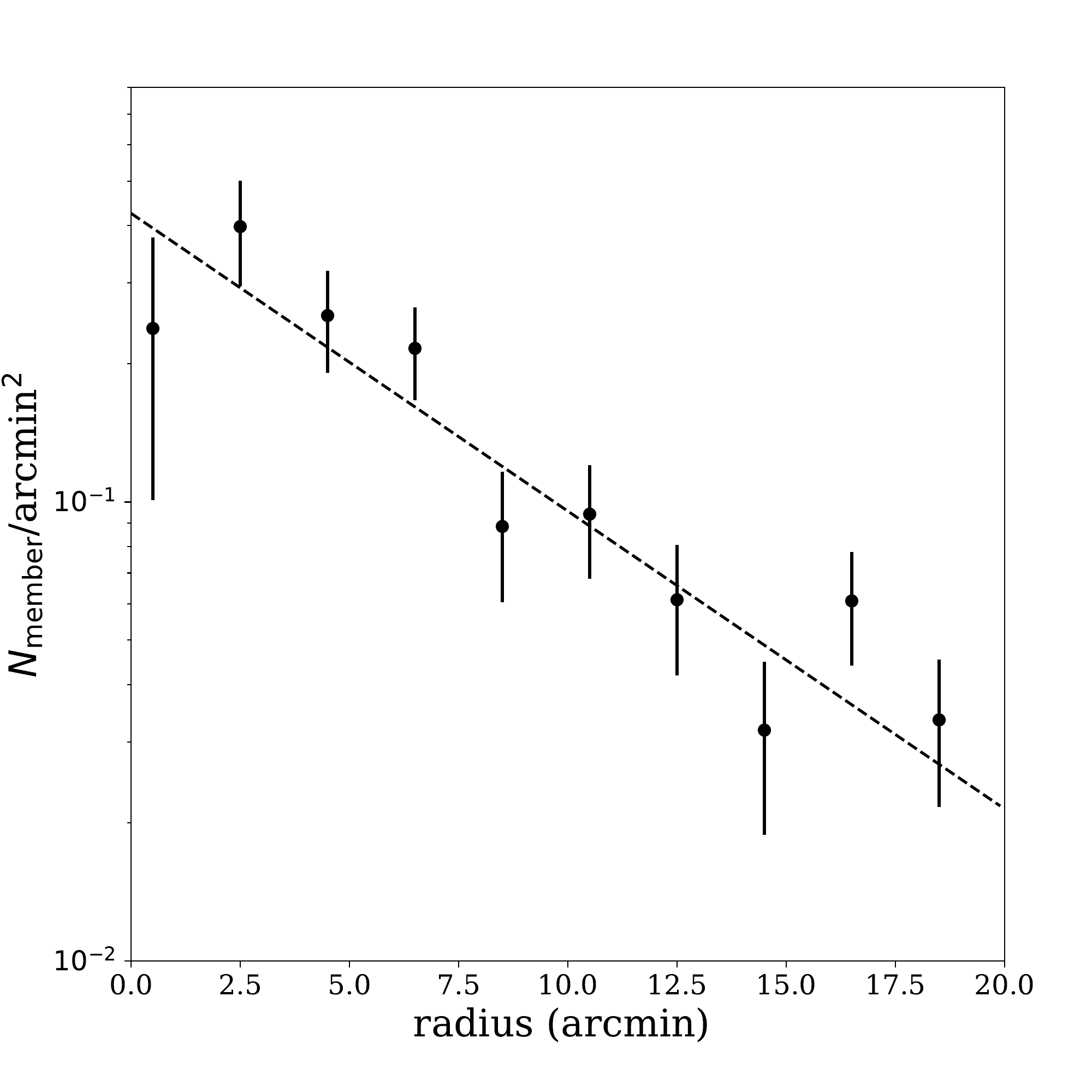}
    \caption{Radial density profile of And~XIX using only probable members. The dashed line is a fitted exponential profile, which results in a scale radius of $r_{\rm half}=11.2\pm2.6^{\prime}$, consistent with that found by \citet{martin16c}.}
    \label{fig:kprof}
\end{figure}

Given the proximity of this stream, there is a possibility that our And~XIX imaging is contaminated with stars from this feature, which could lead us to over-estimate the size of And~XIX. As a check, we measure the half-light radius of And~XIX using only our probable member stars (with $P_{\rm member}>0.1$).  As in \citet{martin16c}, we assume that the stellar density profile of And~XIX can be described using an exponential profile, such that:

\begin{equation}
\rho_{\rm dwarf}=\frac{1.68}{2\pi r_{\rm half}^2(1-\epsilon)}N^*\exp{(-1.68r/r_{\rm half})},
\end{equation}

\noindent where $\epsilon=0.58$ is the ellipticity of And~XIX as measured by \citet{martin16c}, and $N^*$ is the number of stars within the system (in this case, our 81 members). $r$ is the elliptical radius of a given star, which is related to the projected sky coordinates ($x,y$) such that:

\begin{multline}
r=\Bigl(\left[\frac{1}{1-\epsilon}((x-x_0)\cos{\theta}-(y-y_0)\sin{\theta})\right]^2\\
+[(x-x_0)\sin{\theta}+(y-y_0)\cos{\theta}]^2)^{1/2},
\end{multline}

\noindent with the position angle of the major axis of $\theta=34\pm5$, and the central coordinates as derived in \citet{martin16c}. We use the least-squares SciPy.optimize.curve\_fit  routine to derive a half light radius for And~XIX using these equations \citep{scipy}. Our radial profile is shown in fig.~\ref{fig:kprof}, with the best fit profile shown as a dashed line. Using only our kinematic members, we derive $r_{\rm half}=11.2\pm2.6$~arcmin ($2.7\pm0.6$~kpc), which is perfectly consistent with the \citet{martin16c} value of $r_{\rm half}=14.2^{+3.4}_{-1.9}$~arcmin ($3.1^{+0.9}_{-1.1}$~kpc). Our uncertainty only includes that of the fit, whereas the \citet{martin16c} values also marginalise over all other fitted structural properties of And~XIX, and the uncertainty in the distance to And~XIX.

\section{Discussion on the nature of And~XIX}
\label{sec:disc}

\subsection{And~XIX - born this way, or a tidally puffed-up dwarf galaxy?}
\label{sec:veldisc}

With its extended, diffuse appearance and low dark matter density, And~XIX is almost unique in its properties. The only similar galaxy discovered to date is Ant~II \citep{torrealba18}. Ant~II is $\sim2.5$ times fainter than And~XIX ($L=3.4\times10^5~L_\odot$ cf. $L=8.6\times10^5~L_\odot$), and hence lower in surface brightness ($\mu_0=31.9$~mag per sq. arcsec cf. $\mu_0=29.3$~mag per sq. arcsec), but otherwise incredibly similar to And~XIX. The half-light radii, velocity dispersions, central dark matter masses and densities of these two galaxies are all consistent within $1\sigma$, as can be seen in figures~\ref{fig:mrh} and \ref{fig:vmax}. Their mass-to-light ratios are also similarly offset from their expected values as shown in fig.~\ref{fig:mll}. Could there be a common origin for these unique systems?

Using detailed modelling, \citet{torrealba18} showed that the extremely low density halo of Ant~II could be consistent with a cored dark matter halo that has undergone significant tidal stripping. Using the proper motions of Ant~II, \citet{erkal19b} demonstrated that it comes within 26~kpc of the Milky Way, making it a likely candidate for tidal stripping and harassment. Given the similarities between And~XIX and Ant~II, it seems likely that it too, could be a tidally disturbed satellite.

We consider the evidence for And~XIX being a dwarf galaxy that has been ``puffed-up'' by its interactions with M31. First, we turn to our imaging data. Our view of And~XIX from PAndAS is shown in fig.~\ref{fig:a19pandas}. We have ruled out an association with the stream to the west of the galaxy, due to to its higher metallicity. However, the immediate vicinity of the dwarf galaxy still shows signs of disequilibria. While the centre of And~XIX appears smooth and elliptical, the outskirts appear much more distorted, with a potential tidal extension to the north, and some signs of tidally shocked debris to the east and west (similar to what is seen in Hercules in the Milky Way, \citealt{roderick15,kuepper17}). While these outer regions are inherently low surface brightness, they are detected at a similar significance ($>3\sigma$ above the background, \citealt{martin13b}), to a number of confirmed outer halo substructures in M31. These include the South West cloud \citep{bate14} and the nearby stream we investigate in this work. As such, they are indicative of potential tidal stripping and/or shocking in And~XIX. 
 
Additionally, we find evidence for a marginal velocity gradient along the major axis of And~XIX of \vg (equivalent to $\vgrad = -2.1\pm1.8~\kms$kpc$^{-1}$). Whether this is a real rotation signal, or caused by the unusually cold clump we find along the major axis is not clear, but here we discuss the potential causes for this velocity gradient. There are two mechanisms that could give rise to a velocity gradient in And~XIX. The first is ordinary intrinsic rotation. If dSph galaxies originally form as disky systems, they will be tidally `stirred' as they orbit a more massive galaxy. In this scenario, one would expect the dSph to retain some of its initial rotation  \citep{kazantzidis11}. Given that And~XIX's current velocity gradient appears low compared to its dispersion, it could be on an orbit around Andromeda that has erased most of its intrinsic rotation. Such behaviour is seen in the simulations of \citet{kazantzidis13}, where initially disky dwarfs embedded in low density (cored) dark matter halos have their rotational velocities gradually erased by tidal shocks that occur at pericentre passages. At the extreme, such tidal stripping and shocking processes can also lead to a tidally-induced velocity gradient along the direction of elongation (i.e. its major axis, \citealt{martin10}), as the dwarf galaxy is pulled into a stream. The marginal gradient we detect lies along the major axis of And~XIX, but that is expected for  either residual rotation, or tidal streaming motions. As such, we cannot conclude that this gradient is the result of tidal processes.

If the velocity gradient is from normal rotation, and the distorted outer isophotes are instead noise in our images, it could be that And~XIX merely formed with its diffuse extended appearance. However, this would make it an extreme outlier in size, given its brightness. Based on the size-luminosity relation derived using the GAMA survey \citep{lange15}, a galaxy with the current luminosity of And~XIX should have an effective radius of $r_{\rm eff}=0.67$~kpc (with typically 10\% scatter). And~XIX is roughly 5 times this size, making it a clear outlier to observations. 

Turning to simulations, there have yet to be any that successfully produce galaxies with the size and stellar mass of And~XIX through processes such as high-spin halos, or pure feedback (e.g. \citealt{amorisco16a, rong17, chan18}). Further,  even when combining feedback with tidal ``puffing'', the NIHAO simulations \citep{jiang19} fail to produce galaxies with sizes and stellar masses comparable to And~XIX. Galaxies with half-light radii of $\sim2-3$~kpc are typically at least an order of magnitude brighter than And~XIX. The only simulations that may be capable of producing systems like And~XIX (and Ant~II) are those that appeal to tidal processes. Recent works by \citet{ogiya18,carleton18} and \citet{amorisco19a} have demonstrated that tidal heating and stripping of low-density dwarf galaxies can both increase their effective radii, and lower their central dark matter densities. This results in a lower measured central mass for a given effective radius, just as we see for And~XIX. These simulations require galaxies And~XIX to either be on an extreme orbit, to have begun in a low density (cored or low concentration) halo, or possibly both.

Based on a combination of our imaging and spectroscopic data,  a tidal shocking and heating scenario could explain the properties of And~XIX. Such a process could account for the galaxy's extended, diffuse nature, its velocity gradient along its major axis (aligning with potential tidal features seen in the PAndAS imaging), its elevated mass-to-light ratio, and the distorted stellar populations in the PAndAS imaging. There is also the possibility of dynamically cold substructure in the north of the galaxy, where we see the velocity dispersion is lower than the bulk of the galaxy. What this could be (a stream, a sign of a merger) is unclear with our current dataset. On balance, we favour the interpretation that And~XIX is a tidally puffed galaxy, but this would need to be confirmed with detailed modelling of its dynamics, similar to what has been undertaken for Ant~II.

\subsection{And~XIX as a local UDG analogue}
\label{sec:udg}

As seen in fig.~\ref{fig:rhmu}, And~XIX sits at an extreme position within the $r_{\rm half}$ vs. $\mu_0$ plane. While it is consistent with the loose formal definition of a UDG ($r_{\rm half}>1.5$~kpc, $\mu_0>24$~mag/sq. arcsec), it is currently distinct from the canonical UDG population. Can we therefore consider it to be a low luminosity counterpart to the UDG population (which are typically a few orders of magnitude more luminous), or is it something altogether different? To assess this, we should move beyond simple comparisons of structural properties, and compare And~XIX with UDGs for which spectroscopic data exist also. Some groups have measured velocity dispersions (and hence, halo masses) for a selection of UDGs in the Coma \citep{vandokkum19b} and Virgo clusters \citep{toloba18}, as well as two objects in proximity to NGC-1052 \citep{emsellem19a,danieli19a,danieli19b,vandokkum19a}. The majority of these systems have high velocity dispersions and mass-to-light ratios, seemingly in excess of what would be expected from their diffuse stellar populations. The exceptions to this are the UDGs around NGC 1052 (NGC 1052-DF2 and -DF4), whose low velocity dispersions imply a much lower mass-to-light ratio than other UDGs. This is evident in fig.~\ref{fig:mll}, where both NGC 1052-DF2 and DF4 are located in the lower right corner with $[M/L]_{\rm half}\lessapprox2$ (with DF2's exact location dependent on a confirmation of its distance, which is still debated \citealt{trujillo18,vandokkum18b,danieli19b}). While distinct from other UDGs, they have a similar central mass-to-light ratio as several local dwarfs including IC1613 ($r_{\rm half}=1040$~pc, $[M/L]_{\rm half}=2.2\pm0.5~M_\odot/L_\odot$ \citealt{kirby17}), and Sagittarius ($r_{\rm half}=1550$~pc, $[M/L]_{\rm half}=2.2\pm0.2~M_\odot/L_\odot$, \citealt{walker09a}). If NGC 1052-DF2's distance is overestimated, it would be much more similar to field dwarf galaxies. In either event, neither DF2 nor DF4 resemble And~XIX, nor the majority of the UDG population dynamically, so we neglect them in the discussion below.

If we compare And~XIX to the remainder of its more luminous UDG counterparts, we see from fig.~\ref{fig:mrh} that it undoubtedly sits in a lower mass halo. But, with a stellar mass that is approximately 2-3 orders of magnitude lower than the Coma and Virgo UDGs, this is unsurprising. However, if we consider the mass-to-light ratios of And~XIX and UDGs, we begin to see similarities beyond the extended radii and low surface brightness. In fig.~\ref{fig:mll}, we see that And~XIX has a $[M/L]_{\rm half}$ that is $\sim15\times$ that expected for its luminosity (dashed line). This may merely imply that UDGs and And~XIX are typically dark matter dominated systems that we are sampling a larger radius in the dark matter halo than we would for similarly luminous, yet more compact, galaxies. Interestingly, UDGs are also offset from this relation by the same amount. In fact, And~XIX, along with the UDGs and a few Milky Way dSphs (labelled) seem to form a second, offset population in this parameter space. This could point to a common origin for both the more massive, distant UDGs, and And~XIX. The same is true for two of the Milky Way dSphs which show elevated mass-to-light ratios for their luminosity. Triangulum II (Tri~II) is a recently discovered object which shows some evidence for a flaring velocity dispersion profile as a function of radius (\citealt{martin16b}, although see also \citealt{kirby17b}). Such behaviour is expected for dwarf galaxies which are being tidally disrupted. Similarly, some have suggested that Ursa Major (UMa~I) is also being disrupted, as deep imaging shows it to be elongated and irregular in shape \citep{okamoto08}. In addition, one of the UDGs in the Virgo cluster (VLSB-D, labelled in fig.~\ref{fig:mll}) shows significant evidence for tidal disruption in both its structural properties and dynamics \citep{toloba18}, and it too is offset in this parameter space.

As discussed above, the kinematics of And~XIX are complex. While there are signs of a velocity gradient along the major axis of the system, it is unclear whether this is the result of tidal forces, substructure within the galaxy, or ordinary rotation. As such, simple mass estimators that link the velocity dispersion of a galaxy to its enclosed mass may not be appropriate as they assume no significant rotation, and dynamical equilibrium. Both these assumptions may not be appropriate for And~XIX. 

Finally, we note that not all tidally affected dwarf galaxies show high mass-to-light ratios. The Sagittarius dwarf, which is known to be tidally disrupting around the Galaxy \citep{ibata01a}, has a relatively low (current) mass-to-light ratio for its stellar mass and size (although this value depends sensitively on how on how the current stellar and dynamical masses are measured). We also note that, while the UDGs of Coma show no obvious tidal tails or distortions in their stellar populations that would suggest they are undergoing extreme dissolution \citep{mowla17}, this is not necessarily evidence that they have been unaffected by tidal processes. Work by \citet{read06b} and \citet{penarrubia09} have previously shown that tidal tails and streams are transient features, which are typically only visible while the galaxy is close to the pericentre of its orbit. Given the range of properties seen in UDGs, it is likely that there are multiple formation channels for this population. Based on this work, however, it seems plausible that tidal and environmental processes could be an important mechanism, especially in dense environments.

\section{Conclusions}
\label{sec:conc}

In this work, we re-derive the dynamical properties of the unusually diffuse Andromeda satellite, And~XIX, from a sample of $\sim100$ member stars. We summarise our main findings below:

\begin{itemize}
\item We measure a systemic velocity for And~XIX consistent with our previous work, with \vrav, and a velocity dispersion of \sigav, which is higher than the value reported in \citet{collins13}, due to the spatial targeting of the fields selected by those authors. Our analysis also favours a marginal velocity gradient in And~XIX of \vg ($\vgrad = -2.1\pm1.7~\kms$kpc$^{-1}$). 

\item When investigating the kinematics of And~XIX along its projected major axis, we see signs of disequilibria, with the northern-most stars showing a much lower (colder) velocity dispersion than the main body of the dSph. 

\item Assuming And XIX is in dynamical equilibrium, it has an elevated mass-to-light ratio, implying that it is dark matter dominated. Its central mass is low when compared to expectation for galaxies of a similar effective size. However its central mass measurement does place it in a dark matter halo consistent with those of similarly luminous, yet more compact dwarf galaxies. This suggests And~XIX may be a ``puffed up'' dwarf, which has expanded as a result of tidal interactions with its host galaxy.

\item We measure the dynamics and metallicity of a stream feature to the west on And~XIX, which has been suggested to be associated to the dwarf. We measure a systemic velocity of $\langle v_r\rangle=-279.2\pm3.6\kms$, and a dispersion of $\sigma_v=13.8^{+3.5}_{-2.6}~\kms$. Its metallicity is [Fe/H]$=-1.61\pm0.02$~dex, more metal rich than And~XIX ([Fe/H]$=-2.07\pm0.02$~dex). As such, we find it is unlikely to be associated with And~XIX.

\item When comparing the dynamical properties of And~XIX with more luminous UDGs, we see that it behaves similarly, with a lower total mass than would be assumed from its size alone, and an inflated mass-to-light ratio compared to more compact galaxies of a similar luminosity. As such, we find it is a low luminosity analogue of distant UDGs.

\item The unusual kinematics and distorted isophotes seen in the PAndAS imaging suggests that And~XIX has undergone significant tidal interactions. This suggests that the effect of tides and environment may be an important mechanism for the formation of UDGs in dense environments.

\end{itemize}

\section*{Acknowledgements}

We thank Marla Geha for providing helpful comments on this manuscript, and for useful discussions with MLMC on all thing dwarf galaxy related when she was based at Yale University, where some of these data were acquired. We also thank Ana Bonaca for her help during observing runs. MLMC acknowledges previous funding from  NASA through Hubble Fellowship grant \#51337 awarded by the Space Telescope Science Institute, which is operated by the Association of Universities for Research in Astronomy, Inc., for NASA, under contract NAS 5-26555. NFM and RI gratefully acknowledges support from the French National Research Agency (ANR) funded project ``Pristine'' (ANR-18-CE31-0017) along with funding from CNRS/INSU through the Programme National Galaxies et Cosmologie. This work has been published under the framework of the IdEx Unistra and benefits from a funding from the state managed by the French National Research Agency as part of the investments for the future program. K.M.G. acknowledges support provided by NSF grant AST-1614569.
This work makes use of the \texttt{numpy, scipy, matplotlib} and \texttt{astropy} python packages \citep{numpy,scipy,matplotlib,astropy}. 
The data presented herein were obtained at the W. M. Keck Observatory, which is operated as a scientific partnership among the California Institute of Technology, the University of California and the National Aeronautics and Space Administration. The Observatory was made possible by the generous financial support of the W. M. Keck Foundation. Also, based in part on data collected at Subaru Telescope, which is operated by the National Astronomical Observatory of Japan. The authors wish to recognize and acknowledge the very significant cultural role and reverence that the summit of Maunakea has always had within the indigenous Hawaiian community.  We are most fortunate to have the opportunity to conduct observations from this mountain.


\bibliographystyle{mnras}
\bibliography{michelle} 


\appendix

\section{Details of And~XIX member stars}

Here, we present a table of all the probable members of And~XIX (e.g. with $P_{\rm member}>0.1$) in table~\ref{tab:allstars}. We provide a similar catalogue for all 627 of our observed stars electronically with this article, for those interested.

\begin{table*}
	\centering
	\caption{Properties of all member stars for And~XIX. Values of [Fe/H]$=9.99$ indicate a failure to determine a metallicity for that object}
	\label{tab:allstars}
\begin{tabular}{cccccccc}
\hline
RA (deg) & Dec (deg) & $V$-mag & $I$-mag & $v_{\rm helio}$ & $S/N$ (\AA$^{-1}$) & [Fe/H] (dex) & $P_{\rm member}$ \\
\hline
4.888667 & 35.013222 & 21.74 & 20.53 & -84.0 $\pm$ 4.3 & 13.748 & -1.2 $\pm$ 0.1 & 0.483 \\
4.904833 & 35.031472 & 22.56 & 21.49 & -121.7 $\pm$ 5.9 & 6.692 & -2.3 $\pm$ 0.7 & 0.593 \\
4.844542 & 35.037556 & 22.33 & 21.17 & -106.4 $\pm$ 5.3 & 9.548 & -1.7 $\pm$ 0.2 & 0.782 \\
4.878042 & 35.046 & 22.27 & 21.04 & -108.6 $\pm$ 4.3 & 11.48 & -1.0 $\pm$ 0.1 & 0.914 \\
4.866375 & 35.049167 & 22.58 & 21.52 & -125.7 $\pm$ 6.1 & 8.96 & -2.5 $\pm$ 0.5 & 0.781 \\
4.955708 & 35.086056 & 22.56 & 21.44 & -108.1 $\pm$ 6.2 & 7.98 & -1.7 $\pm$ 0.2 & 0.914 \\
4.92075 & 35.088333 & 22.48 & 21.33 & -115.1 $\pm$ 4.7 & 9.1 & -1.5 $\pm$ 0.3 & 0.882 \\
4.914333 & 35.089889 & 22.48 & 21.41 & -119.8 $\pm$ 5.1 & 9.128 & -1.6 $\pm$ 0.4 & 0.612 \\
4.886875 & 35.093556 & 22.77 & 21.63 & -100.7 $\pm$ 5.0 & 7.336 & 9.99 $\pm$ 9.99 & 0.611 \\
4.872667 & 35.098028 & 22.73 & 21.57 & -117.1 $\pm$ 7.5 & 7.532 & 9.99 $\pm$ 9.99 & 0.472 \\
4.969583 & 35.099917 & 22.76 & 21.7 & -121.2 $\pm$ 7.1 & 6.608 & 9.99 $\pm$ 9.99 & 0.694 \\
4.910708 & 35.104972 & 22.18 & 20.99 & -102.0 $\pm$ 3.9 & 12.712 & -1.8 $\pm$ 0.3 & 0.665 \\
4.89725 & 35.110278 & 22.0 & 20.73 & -115.9 $\pm$ 3.2 & 14.084 & -1.8 $\pm$ 0.7 & 0.639 \\
4.95925 & 35.122222 & 22.78 & 21.71 & -123.3 $\pm$ 5.6 & 7.392 & 9.99 $\pm$ 9.99 & 0.79 \\
4.964958 & 35.125056 & 22.42 & 21.33 & -114.3 $\pm$ 4.8 & 9.576 & 9.99 $\pm$ 9.99 & 0.642 \\
4.866125 & 35.052139 & 23.09 & 22.02 & -133.0 $\pm$ 8.9 & 4.312 & -0.8 $\pm$ 0.5 & 0.846 \\
4.930958 & 35.121583 & 23.12 & 22.13 & -112.0 $\pm$ 7.3 & 4.592 & 9.99 $\pm$ 9.99 & 0.926 \\
4.959917 & 35.1565 & 23.38 & 22.38 & -126.5 $\pm$ 13.8 & 3.276 & 9.99 $\pm$ 9.99 & 0.717 \\
4.843542 & 34.980056 & 22.28 & 21.1 & -91.6 $\pm$ 4.5 & 12.348 & -1.2 $\pm$ 0.6 & 0.72 \\
4.860083 & 34.985167 & 22.83 & 21.73 & -115.6 $\pm$ 10.3 & 3.164 & 9.99 $\pm$ 9.99 & 0.66 \\
4.816917 & 34.989806 & 22.53 & 21.38 & -96.8 $\pm$ 5.4 & 8.96 & 9.99 $\pm$ 9.99 & 0.867 \\
4.802083 & 35.003722 & 22.46 & 21.23 & -95.6 $\pm$ 4.5 & 9.576 & -2.3 $\pm$ 0.6 & 0.641 \\
4.792917 & 34.949139 & 23.44 & 22.46 & -142.0 $\pm$ 10.4 & 2.324 & 0.0 $\pm$ 0.0 & 0.503 \\
4.824 & 34.995222 & 22.92 & 21.83 & -105.0 $\pm$ 8.6 & 4.984 & -1.1 $\pm$ 0.4 & 0.704 \\
4.844542 & 35.037583 & 22.33 & 21.17 & -107.1 $\pm$ 5.4 & 6.076 & -2.3 $\pm$ 0.2 & 0.782 \\
4.878042 & 35.046056 & 22.27 & 21.04 & -108.2 $\pm$ 5.0 & 8.596 & -1.5 $\pm$ 0.1 & 0.914 \\
4.916542 & 35.068722 & 22.36 & 21.31 & -87.2 $\pm$ 8.0 & 6.3 & -1.5 $\pm$ 0.2 & 0.212 \\
5.005083 & 35.070528 & 21.97 & 20.66 & -114.5 $\pm$ 3.9 & 11.48 & -1.9 $\pm$ 0.1 & 0.692 \\
4.976 & 35.082556 & 22.44 & 21.31 & -103.8 $\pm$ 5.3 & 6.888 & -1.5 $\pm$ 0.2 & 0.725 \\
4.955667 & 35.086 & 22.56 & 21.44 & -116.1 $\pm$ 6.7 & 6.272 & -1.4 $\pm$ 0.2 & 0.914 \\
4.914292 & 35.089889 & 22.48 & 21.41 & -97.7 $\pm$ 5.6 & 5.46 & -2.4 $\pm$ 0.5 & 0.611 \\
4.89725 & 35.11025 & 22.0 & 20.73 & -120.0 $\pm$ 4.2 & 11.172 & -1.8 $\pm$ 0.1 & 0.638 \\
5.0275 & 35.068917 & 21.95 & 20.68 & -111.6 $\pm$ 3.6 & 11.844 & -2.7 $\pm$ 0.2 & 0.235 \\
5.010542 & 35.058111 & 22.95 & 21.98 & -132.9 $\pm$ 11.2 & 2.8 & 9.99 $\pm$ 9.99 & 0.584 \\
4.925542 & 35.067889 & 22.76 & 21.68 & -108.7 $\pm$ 6.9 & 4.004 & -0.9 $\pm$ 0.5 & 0.941 \\
4.923833 & 35.091472 & 23.22 & 22.25 & -117.7 $\pm$ 11.1 & 2.548 & -1.7 $\pm$ 0.3 & 0.886 \\
4.969542 & 35.099889 & 22.76 & 21.7 & -127.1 $\pm$ 10.7 & 5.18 & -1.2 $\pm$ 0.5 & 0.692 \\
4.82325 & 35.106833 & 23.09 & 22.06 & -109.9 $\pm$ 12.7 & 0.952 & 9.99 $\pm$ 9.99 & 0.674 \\
4.920708 & 35.088306 & 22.48 & 21.33 & -121.6 $\pm$ 8.1 & 6.132 & -0.4 $\pm$ 0.1 & 0.881 \\
4.840583 & 35.053444 & 22.37 & 21.2 & -100.8 $\pm$ 7.9 & 6.328 & 0.0 $\pm$ 0.0 & 0.828 \\
4.778875 & 35.064778 & 22.54 & 21.39 & -111.5 $\pm$ 6.1 & 5.684 & 9.99 $\pm$ 9.99 & 0.874 \\
4.753417 & 35.041861 & 22.16 & 20.92 & -100.5 $\pm$ 3.4 & 8.764 & -0.6 $\pm$ 0.1 & 0.81 \\
4.854917 & 35.142889 & 23.24 & 22.18 & -112.8 $\pm$ 10.3 & 2.856 & 9.99 $\pm$ 9.99 & 0.565 \\
4.941708 & 35.16325 & 22.79 & 21.78 & -119.3 $\pm$ 10.2 & 3.416 & -0.9 $\pm$ 0.5 & 0.593 \\
4.873792 & 35.186583 & 23.01 & 21.95 & -128.8 $\pm$ 13.5 & 2.744 & -0.7 $\pm$ 0.5 & 0.557 \\
4.993417 & 35.075889 & 22.39 & 21.21 & -112.2 $\pm$ 6.0 & 5.124 & 9.99 $\pm$ 9.99 & 0.867 \\
5.016125 & 35.124694 & 22.09 & 20.85 & -112.3 $\pm$ 4.8 & 11.144 & -2.2 $\pm$ 0.1 & 0.659 \\
5.017958 & 35.134917 & 22.1 & 20.88 & -103.5 $\pm$ 4.4 & 10.808 & -2.5 $\pm$ 0.1 & 0.537 \\
5.017417 & 35.146639 & 22.61 & 21.56 & -86.4 $\pm$ 10.5 & 4.76 & -2.0 $\pm$ 0.4 & 0.644 \\
4.843542 & 34.980028 & 22.28 & 21.1 & -102.6 $\pm$ 5.0 & 7.28 & -1.1 $\pm$ 0.1 & 0.728 \\
4.866375 & 35.049194 & 22.58 & 21.52 & -108.6 $\pm$ 8.2 & 3.92 & -1.6 $\pm$ 0.5 & 0.784 \\
4.701333 & 34.843167 & 21.62 & 20.35 & -110.9 $\pm$ 4.3 & 11.508 & 9.99 $\pm$ 9.99 & 0.47 \\
4.95775 & 34.997389 & 22.17 & 20.86 & -98.5 $\pm$ 4.3 & 10.248 & -2.1 $\pm$ 0.1 & 0.853 \\
4.955333 & 35.061306 & 22.42 & 21.29 & -107.6 $\pm$ 7.6 & 6.86 & -2.9 $\pm$ 0.3 & 0.643 \\
5.035667 & 34.95425 & 22.55 & 21.46 & -125.9 $\pm$ 6.8 & 7.0 & -0.2 $\pm$ 0.5 & 0.73 \\
5.01625 & 34.980417 & 21.96 & 20.55 & -128.3 $\pm$ 3.8 & 13.356 & -1.9 $\pm$ 0.1 & 0.774 \\
5.092375 & 34.887972 & 22.58 & 21.34 & -101.8 $\pm$ 5.6 & 6.608 & -2.2 $\pm$ 0.2 & 0.143 \\
4.979625 & 34.979417 & 22.92 & 21.91 & -104.5 $\pm$ 8.9 & 4.2 & 9.99 $\pm$ 9.99 & 0.686 \\
5.014208 & 34.991611 & 23.04 & 21.97 & -114.4 $\pm$ 9.5 & 3.5 & -0.9 $\pm$ 0.2 & 0.656 \\
5.09975 & 34.888667 & 23.24 & 22.19 & -106.8 $\pm$ 12.6 & 3.08 & -0.0 $\pm$ 0.5 & 0.364 \\
4.886875 & 35.093528 & 22.77 & 21.63 & -108.4 $\pm$ 6.2 & 5.46 & -2.5 $\pm$ 0.5 & 0.611 \\
4.850708 & 35.121528 & 22.79 & 21.63 & -117.3 $\pm$ 5.8 & 5.04 & -2.2 $\pm$ 0.5 & 0.418 \\
4.9035 & 35.181917 & 22.97 & 22.05 & -110.5 $\pm$ 13.6 & 2.744 & -1.6 $\pm$ 0.4 & 0.228 \\
\end{tabular}
\end{table*}

\begin{table*}
 \contcaption{}
 \label{tab:continued}
 \begin{tabular}{cccccccc}
\hline
RA (deg) & Dec (deg) & $V$-mag & $I$-mag & $v_{\rm helio}$ & $S/N$ (\AA$^{-1}$) & [Fe/H] (dex) & $P_{\rm member}$ \\
\hline
4.877625 & 35.246194 & 22.93 & 21.79 & -120.4 $\pm$ 8.7 & 3.36 & 9.99 $\pm$ 9.99 & 0.352 \\
4.822417 & 34.937917 & 22.43 & 21.25 & -96.7 $\pm$ 7.5 & 10.444 & -2.1 $\pm$ 1.0 & 0.772 \\
4.825625 & 34.938333 & 22.29 & 21.1 & -126.4 $\pm$ 8.6 & 11.732 & -2.1 $\pm$ 0.6 & 0.74 \\
4.873167 & 34.95125 & 22.19 & 20.98 & -121.7 $\pm$ 5.1 & 12.32 & 9.99 $\pm$ 9.99 & 0.783 \\
4.833583 & 34.963444 & 22.14 & 20.94 & -118.1 $\pm$ 8.4 & 12.768 & 9.99 $\pm$ 9.99 & 0.627 \\
4.869292 & 34.975667 & 22.67 & 21.58 & -136.3 $\pm$ 14.9 & 7.196 & 9.99 $\pm$ 9.99 & 0.809 \\
4.962625 & 35.023333 & 22.63 & 21.56 & -104.2 $\pm$ 5.9 & 6.104 & -1.5 $\pm$ 0.3 & 0.814 \\
4.955375 & 35.06125 & 22.42 & 21.29 & -88.8 $\pm$ 9.1 & 8.596 & -0.1 $\pm$ 0.2 & 0.635 \\
4.916542 & 35.068694 & 22.36 & 21.31 & -59.2 $\pm$ 13.5 & 8.596 & 9.99 $\pm$ 9.99 & 0.103 \\
4.802792 & 34.894972 & 22.9 & 21.82 & -107.4 $\pm$ 9.6 & 6.3 & -1.6 $\pm$ 0.4 & 0.651 \\
4.888 & 34.94725 & 23.27 & 22.21 & -94.1 $\pm$ 5.5 & 4.34 & 9.99 $\pm$ 9.99 & 0.563 \\
4.967417 & 35.072 & 23.56 & 22.58 & -71.7 $\pm$ 12.5 & 2.66 & 9.99 $\pm$ 9.99 & 0.617 \\
4.971708 & 34.954861 & 22.72 & 21.66 & -104.3 $\pm$ 5.2 & 10.528 & -1.9 $\pm$ 0.2 & 0.848 \\
4.957792 & 34.997361 & 22.17 & 20.86 & -117.2 $\pm$ 2.9 & 20.608 & -1.7 $\pm$ 0.1 & 0.855 \\
5.014167 & 34.991611 & 23.04 & 21.97 & -98.3 $\pm$ 7.3 & 9.464 & -1.1 $\pm$ 0.1 & 0.655 \\
5.0485 & 35.008194 & 22.92 & 21.87 & -108.6 $\pm$ 3.4 & 8.848 & -2.8 $\pm$ 0.3 & 0.768 \\
5.026375 & 35.027056 & 23.1 & 22.04 & -116.2 $\pm$ 10.5 & 7.42 & -2.2 $\pm$ 0.2 & 0.614 \\
4.656625 & 34.982472 & 22.22 & 21.01 & -104.9 $\pm$ 3.4 & 13.3 & -1.3 $\pm$ 0.1 & 0.556 \\
4.709542 & 35.057833 & 22.13 & 21.01 & -115.9 $\pm$ 4.6 & 12.852 & -2.0 $\pm$ 0.1 & 0.144 \\
4.933083 & 35.211583 & 22.22 & 21.12 & -113.3 $\pm$ 6.1 & 11.9 & -1.9 $\pm$ 0.1 & 0.584 \\
\hline
\end{tabular}
\end{table*}

\section{Duplicate spectra and pipeline comparison}
\label{sec:pipeline}
\begin{figure}
     \includegraphics[angle=0,width=\columnwidth]{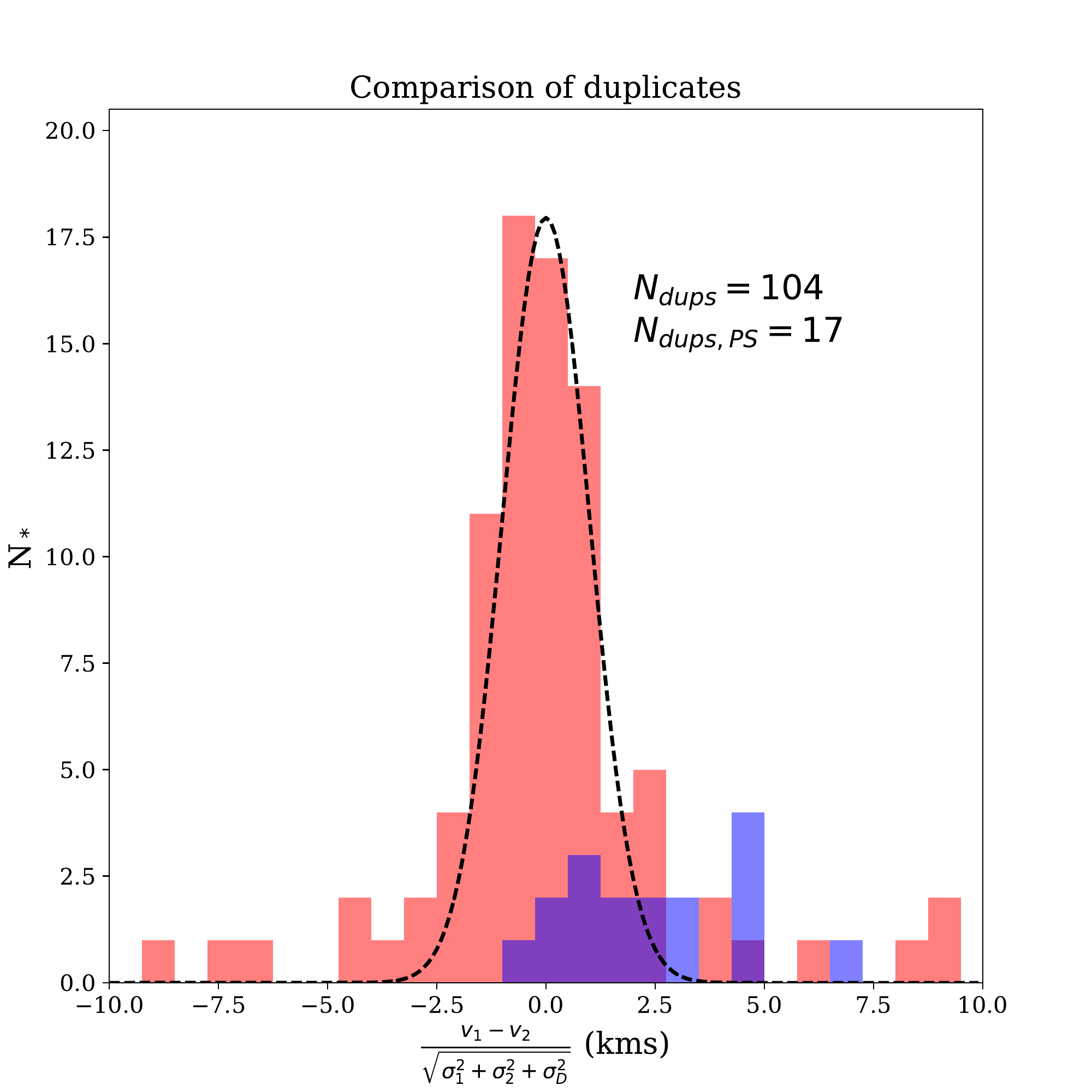}
    \caption{The red histogram is a comparison of the normalised error distribution of 104 stars with duplicated observations. The normalised error incorporates the velocity differences between the repeat measurements ($v_1$ and $v_2$), their uncertainties ($\sigma_1$ and $\sigma_2$), and the systematic uncertainty for the DEIMOS instrument ($\sigma_D = 3.2$ kms$^{-1}$, \citealt{simon07}). The dashed line represents a unit Gaussian ($\mu = 0,\sigma=1$), which well fits our data. The blue histogram is of the 17 stars in common between the PAndAS and SPLASH pipelines. The bulk to the stars follow the same distribution as the main duplicate sample, but there is some evidence of a shift between the two pipelines of order 2.4 $\kms$.} 
    \label{fig:vcomp}
\end{figure}

\begin{figure}
     \includegraphics[angle=0,width=\columnwidth]{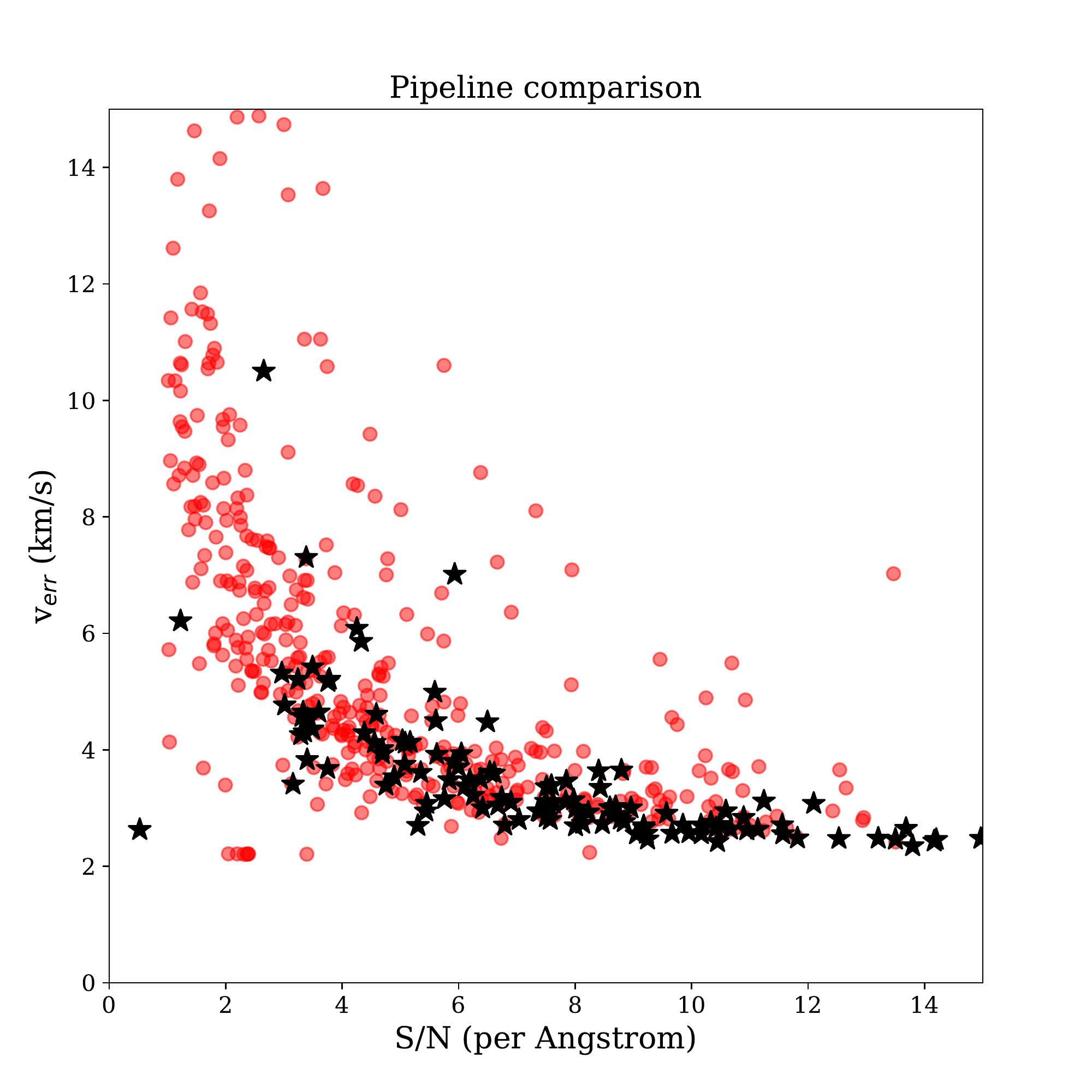}
    \caption{A comparison of the measured velocity uncertainties using the PAndAS pipeline (red circles) and SPLASH pipeline (black stars), as a function of $S/N$. In both cases, we see that the uncertainties increase as the $S/N$ increases, and the trends are virtually identical for both pipelines.  }
    \label{fig:verrpipe}
\end{figure}

Within our dataset, we have 104 stars that were observed with DEIMOS on more than one occasion. This dataset allows us to investigate the reliability of our velocity measurements. We compare the velocities measured for this sample, which allows us to trace the ``true'' error distribution for our sample, akin to the work of e.g. \citet{simon07}, who used their sample of 49 stars with repeat observations to determine a normalised error distribution for the DEIMOS instrument ($\sigma_N$). This normalised error distribution can be thought of as the velocity difference between repeat observations ($v_1-v_2$), normalised by their measured uncertainties ($\sigma_1$ and $\sigma_2$), and a systematic uncertainty from DEIMOS itself ($\sigma_D)$, such that:

\begin{equation}
\sigma_N=\frac{v_1-v_2}{\sqrt{\sigma_1^2 + \sigma_2^2+\sigma_D^2}}
\end{equation}

In their work, \citet{simon07} found $\sigma_D=2.2\kms$. In fig.~\ref{fig:vcomp}, we show our normalised error distribution for 104 duplicate observations, and repeat their analysis to derive $\sigma_D=3.2\kms$  for our dataset. As such, we use this marginally larger value throughout our analysis, adding it in quadrature to the measurement uncertainties for our velocities.

For this study, 3 of our 11 spectroscopic masks were reduced with the SPLASH pipeline (A19l1, A19l2, A19r1), while  the remaining 8 were reduced with PAndAS. In total, we have 17 stars in common between these two datasets. We can compare these duplicates in the same way as above to see if there are any systematic offsets between the two reduction methods. This is shown as the blue histogram in fig.~\ref{fig:vcomp}. The sample size here is much smaller, but we see some evidence for a slight offset between the two pipelines. The mean value for $\sigma_N$ is $2.0\kms$ for this data set, with the SPLASH velocities being slightly higher on average than the PAndAS dataset. This could imply that the two different methods for determining velocities produce slightly inconsistent results. But it is hard to be sure from only 17 stars.

We can further compare these two reduction methods by comparing how their velocity uncertainties degrade with $S/N$ of the observed spectra. Here, we can use the full sample of stars from both pipelines (721 for PAndAS vs. 284 for SPLASH). The results are shown in fig.~\ref{fig:verrpipe}. The red circles show the uncertainties for the PAndAS pipeline, while the black stars show the SPLASH pipeline. Here we see that the behaviour for both pipelines is almost identical. The uncertainties slowly increase towards a $S/N$ of 4\AA$^{-1}$, at which point they increase more rapidly. This consistency implies that the two different methods for measuring the uncertainties in these pipelines (MCMC vs. Monte Carlo) produce very similar outcomes.

\section{Spectra for low $\sigma_v$ stars}
\begin{figure*}[!h]
     \includegraphics[angle=0,width=\columnwidth]{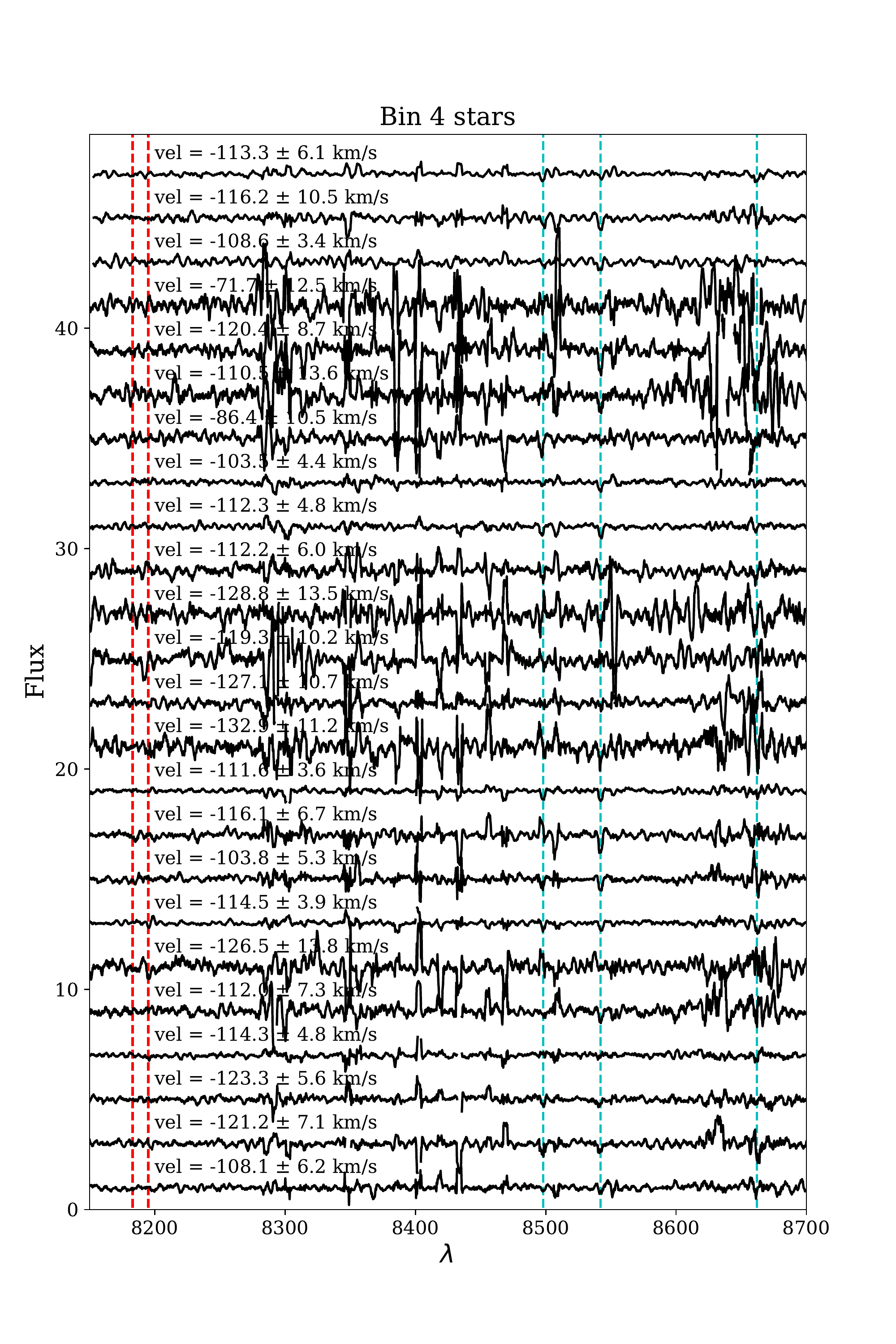}
          \includegraphics[angle=0,width=\columnwidth]{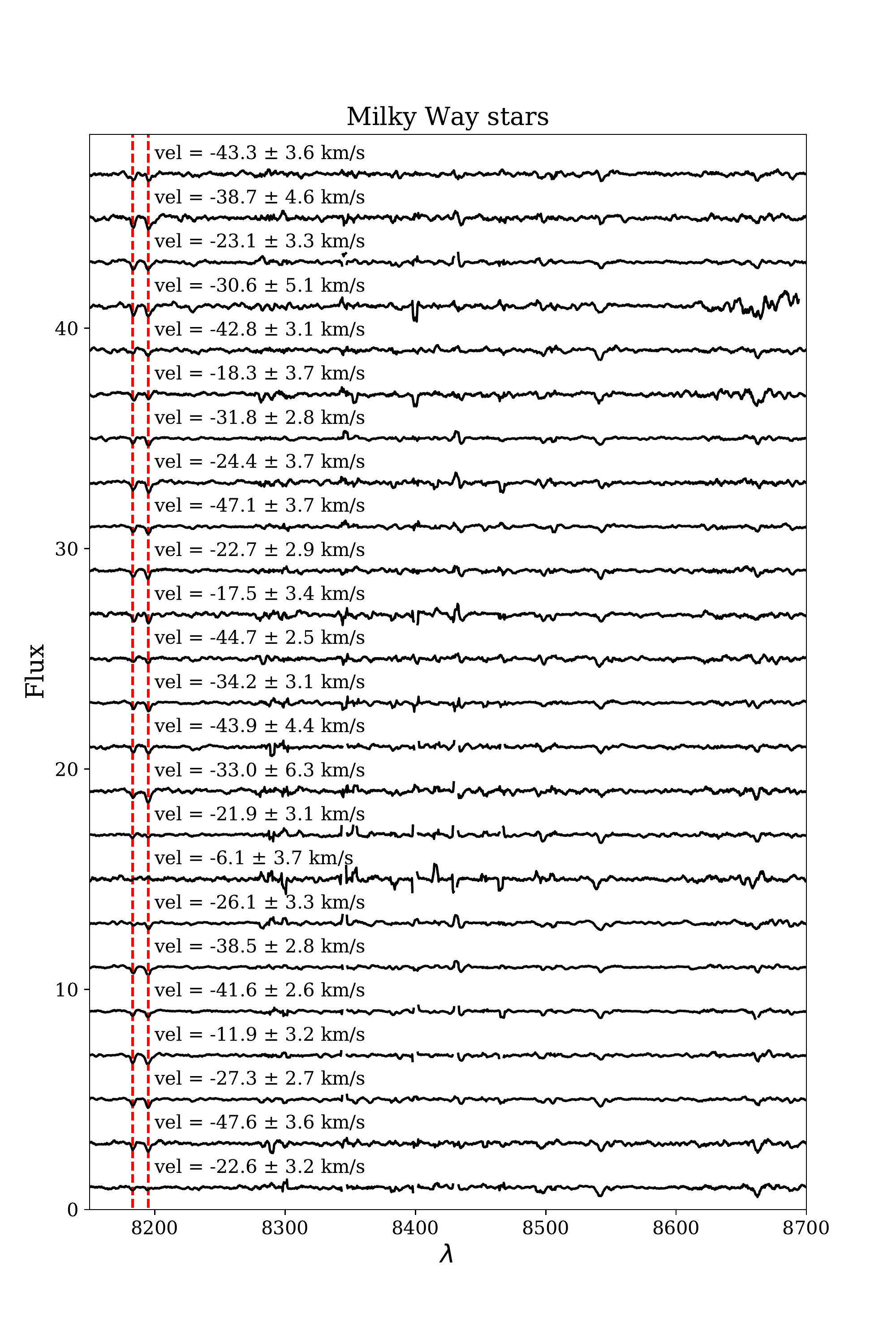}
    \caption{Spectra for all stars in the low velocity dispersion bin along the major axis with $S/N>3$\AA$^{-1}$ (left). There is no evidence for significant absorption at the location of the Na~I doublet (red dashed lines). However, this is more clearly seen in probable Milky Way stars (right, randomly selected examples).}
    \label{fig:bin4}
\end{figure*}

Here we show the spectra for all the stars in the low velocity dispersion bin along the major axis in fig.~\ref{fig:bin4}. The gravity sensitive Na I lines are indicated with vertical red dashed lines. For the member stars, we do not see strong absorption at this location. However, when comparing with 24 randomly selected Milky Way stars, we see clearer absorption. This adds confidence that this low dispersion bin is not the result of including Milky Way contaminants.


\bsp	
\label{lastpage}
\end{document}